\documentclass[iop]{emulateapj}
\usepackage{graphicx,epsfig,epsf,rotating,fancyhdr,amsmath,natbib}
\usepackage{apjfonts}

\newcommand{\kms}{km~s$^{-1}$}

\newcommand{\arc}{{\arcsec}}

\shorttitle{flare-driven coronal rain}

\shortauthors{Scullion et al.}

\begin{document}

\title{Observing the formation of flare-driven coronal rain}

\author{E.~Scullion\altaffilmark{1,2}, L.~Rouppe van der Voort\altaffilmark{1}, P.~Antolin\altaffilmark{3}, S.~Wedemeyer\altaffilmark{1},  G.~Vissers\altaffilmark{1}, \\
E.~P.~Kontar\altaffilmark{4}, P.~Gallagher\altaffilmark{2}}

\altaffiltext{1}{Institute of Theoretical Astrophysics, University of Oslo, P.O. Box 1029, Blindern, NO-0315 Oslo, Norway}
\altaffiltext{2}{Trinity College Dublin, College Green, Dublin 2, Ireland: scullie@tcd.ie}
\altaffiltext{3}{National Astronomical Observatory of Japan, 2-21-1 Osawa, Mitaka, Tokyo 181-8588, Japan}
\altaffiltext{4}{SUPA School of Physics and Astronomy, University of Glasgow, G12 8QQ, UK}

\begin{abstract} 

Flare-driven coronal rain can manifest from rapidly cooled plasma condensations near coronal loop-tops in thermally unstable post-flare arcades. We detect 5 phases that characterise the post-flare decay: heating, evaporation, conductive cooling dominance for $\sim$120~s, radiative / enthalpy cooling dominance for $\sim$4700~s and finally catastrophic cooling occurring within 35-124~s leading to rain strands with s periodicity of 55-70~s. We find an excellent agreement between the observations and model predictions of the dominant cooling timescales and the onset of catastrophic cooling. At the rain formation site we detect co-moving, multi-thermal rain clumps that undergo catastrophic cooling from $\sim$1~MK to $\sim$22000~K. During catastrophic cooling the plasma cools at a maximum rate of 22700~K~s$^{-1}$ in multiple loop-top sources. We calculated the density of the EUV plasma from the DEM of the multi-thermal source employing regularised inversion. Assuming a pressure balance, we estimate the density of the chromospheric component of rain to be 9.21$\times$10$^{11}$~$\pm$1.76$\times$10$^{11}$~cm$^{-3}$ which is comparable with quiescent coronal rain densities. With up to 8 parallel strands in the EUV loop cross section, we calculate the mass loss rate from the post-flare arcade to be as much as 1.98$\times$10$^{12}$~$\pm$4.95$\times$10$^{11}$~g~s$^{-1}$. Finally, we reveal a close proximity between the model predictions of 10$^{5.8}$~K and the observed properties between 10$^{5.9}$~K and 10$^{6.2}$~K, that defines the temperature onset of catastrophic cooling. The close correspondence between the observations and numerical models suggests that indeed acoustic waves (with a sound travel time of 68~s) could play an important role in redistributing energy and sustaining the enthalpy-based radiative cooling.

\end{abstract}

\keywords{Sun:  -- Methods: observational -- Methods: data analysis -- Techniques: image processing -- Techniques: spectroscopic -- Telescopes}

\section{INTRODUCTION}

\par Coronal rain is a transient phenomena within coronal loops and represents a key component of the mass cycling, between the solar chromosphere and corona and occurs frequently in active regions \citep[for e.g.,][and see references therein]{1970PASJ...22..405K,1972SoPh...25..413L,1977SoPh...51...83L,2001SoPh..198..325S,2007A&A...475L..25O,2012ApJ...745..152A,2013ApJ...771L..29F,2014SoPh..289.4117A,2015ApJ...806...81A}. During its formation, coronal rain momentarily constitutes the finest scale substructures of coronal loops, making it important to investigate with respect to understanding the heating of loops, the building blocks of the solar corona. 

\par Many observational studies of active regions indicate a general tendency for cooling \citep[][and references therein]{2011ApJ...736..111T,2012ApJ...753...35V,2015ApJ...807..158F}. Coronal rain flows within a sheath of hot coronal plasma, as intermittent, elongated and cool (chromospheric) condensations, which follow trajectories tracing magnetic arcades and with densities varying between 2$\times$10$^{10}$~cm$^{-3}$ and 2.5$\times$10$^{11}$~cm$^{-3}$. This results in a substantial mass loss per loop of 1-5$\times$10$^{9}$~g~s$^{-1}$ \citep{2015ApJ...806...81A}. As the loop-top plasma cools and condenses narrow, elongated clumps form that have been observed to fall back to the lower solar atmosphere, at speeds greater than $\sim$40~km~s$^{-1}$.  A comprehensive, statistical analysis of the properties of quiescent coronal rain in H$\alpha$ 656.28~nm line scans, using high resolution ground-based observations, was completed by \citet{2012ApJ...745..152A}, who reported average widths and lengths of $\sim$310~km and $\sim$710~km, respectively. Furthermore, the average temperatures of the dark clumps appear below 7000~K, with an average falling speed of $\sim$70~km~s$^{-1}$. On the other hand, flare-driven coronal rain was reported with an apparent constant projected speed of 134~km~s$^{-1}$ and the downward acceleration is generally no more than 80~m~s$^{-2}$ \citep{2014ApJ...780L..28M}. 

\par Active region coronal rain is observed in many other visible and near-IR and EUV channels, such as Ca~{\sc ii}~854.2~nm and He~{\sc ii}~30.4~nm, as well as multiple other Transition Region (TR) emission limes indicating its multi-thermal structure \citep{2015ApJ...806...81A}. Such cooling progression throughout TR temperatures have previously been considered to explain EUV brightness variations \citep{1976ApJ...210..575F,2001SoPh..198..325S,2007A&A...475L..25O,2009ApJ...694.1256T,2011ApJ...734...90W}. 
Cooling has been shown to continue with delays of up to 103~s between adjacent, parallel propagating strands \citep{2001SoPh..198..325S}. 

\par In this present work, we distinguish between the many detailed studies of non-flaring, widespread, active region coronal rain as a relatively weak form of mass condensation (with respect to mass loading and energy input) and, henceforth, we refer to that as quiescent coronal rain. In this study, we are interested in the relatively stronger (i.e. higher density and infrequent) flare-driven coronal rain, which is investigated to a much lesser extent, due to the infrequency of multi-instrumental studies at sufficiently high resolution and unpredictable flares.

\par There have been many studies aimed at understanding cooling processes in post-flare loops \citep[for e.g.,][and references therein]{1975SoPh...43..189M,1978ApJ...220.1137A,1983ApJ...265.1103D,1985ApJ...289..425F,1993SoPh..147..263C,2003ApJ...585.1087F,2005A&A...437..311B,2006SoPh..234...41K,2006ApJ...637..522W,2007ApJ...666.1245W,2012A&A...543A..90R,2012ApJ...746...18R}. Statistical analysis of the cross-sectional widths of flare-driven coronal rain strands, within post-flare loops, have been reported down to the diffraction limit of the most advanced ground-based instruments with strong implications that rain strands may exist at even narrower widths \citep{2014ApJ...797...36S}. Observational correspondences of red-shifted plasma emission (downflow) after a flare, that was initially heated and blue-shifted (upflow) to fill the post-flare arcade loop system via chromospheric evaporation, has been widely reported \citep[for e.g.,][]{2003ApJ...586.1417B,2008ASPC..397..184R,2014ApJ...780L..28M}. 


\begin{figure}[!hb]
\centering
\includegraphics[clip=true,trim=0cm 1cm 0cm 0cm,scale=0.91, angle=0]{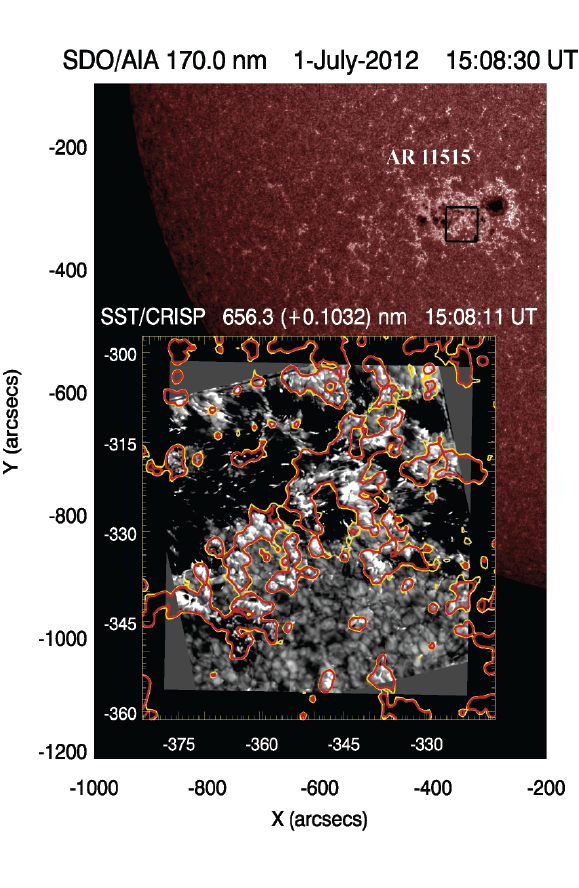}
\caption{The SST/CRISP grey-scale H$\alpha$ 656.28 nm (+ 0.1032 nm) spectral image ({\it inset}) and the SDO/AIA 170.0 nm continuum perspective image ({\it background}), are presented for comparison. The CRISP FOV (white box overlaid on continuum image) is centred on active region (AR) 11515 on 1$^{st}$ July 2012 at 15:08:30~UT. The 170.0~nm photospheric bright points are contoured over the CRISP H$\alpha$ co-incident images and used as a reference for the co-alignment of the datasets in space and time.}
\label{fig1}
\end{figure}



\begin{figure*}[!ht]
\centering
\includegraphics[clip=true,trim=0cm 0cm 0cm 0cm, scale=0.99, angle=0]{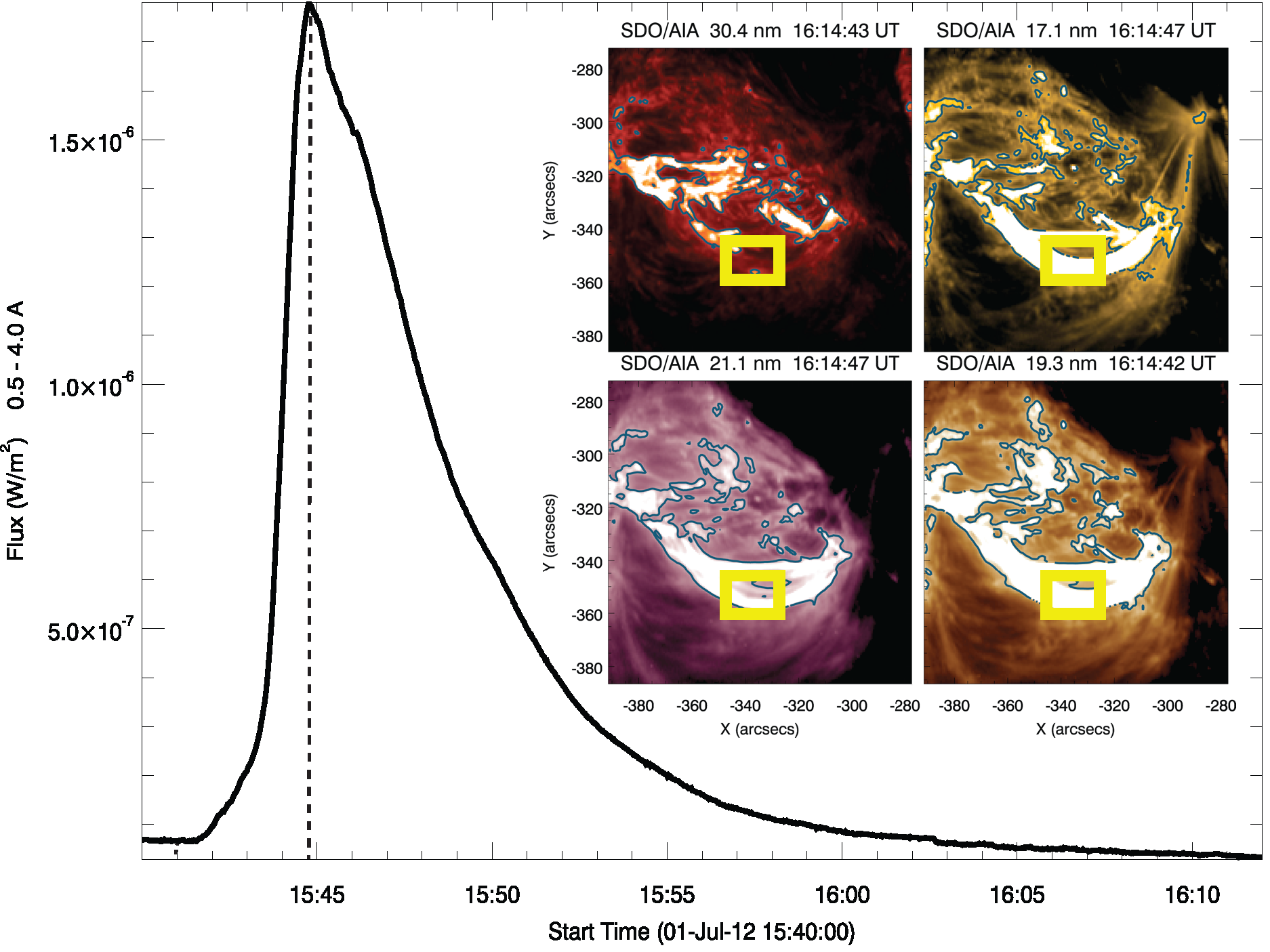}
\caption{{\it Main}: The GOES 3-s X-Ray flux (W/m$^{2}$) light-curve, for the 0.05-0.4~nm passband, is presented in the time range from 15:40~UT until 16:12~UT, for the C8.2-class flare. The vertical dashed line marks the peak time ($\sim$15:44:40~UT) for the flare energy deposition. {\it Inset}: The AIA post-flare EUV loops, as they appear 30~mins after the GOES X-ray flux peak, are presented for He~{\sc ii}~30.4~nm ({\it top left}), Fe~{\sc ix}~17.1~nm ({\it top right}), Fe~{\sc xiv}~21.1~nm ({\it bottom left}) and Fe~{\sc xii}~19.3~nm ({\it bottom right}). The yellow box marks the region of interest within the flaring loop cross-section near the loop-top of the flaring arcade and the focus of the remainder of the coronal rain formation study.}
\label{fig2}
\end{figure*}



\par During the formation of coronal rain, the evaporated plasma arising during flare heating in the impulsive phase, is rapidly cooled via thermal instability below 1-2~MK, Physically, the formation of coronal rain is thought to result from a loop-top thermal instability mechanism \citep{1953ApJ...117..431P,1965ApJ...142..531F,2013ApJ...772...40C} when radiative losses exceed heating input to the coronal loop system. The cooling becomes accelerated at a late stage in this process, known as catastrophic cooling, whereby over-dense hot/warm loops deplete plasma towards their foot-points with progressively faster radiative cooling rates within multi-thermal loop strands. Using model calculations of conductively and radiatively cooled flare plasma \citet{1982ApJ...258..373D} calculated down-flow velocities of 50~km~s$^{-1}$, consistent with quiescent rain observations. Numerical simulations of quiescent coronal rain formation suggests that catastrophic cooling is generally a short-lived and dependent upon the foot-point heating parameters, but it is expected to occur in less than 1 hour typically \citep{2010ApJ...716..154A,2005ApJ...624.1080M,2010ApJ...709..499S}. In the flare-driven scenario, we expect substantially larger foot-point heating coupled with impulsive and intense chromospheric evaporation leading to greater mass loading of post-flare loops systems.

\par Currently very little is understood about the nature of coronal rain at the point of formation in flare arcades, with respect to the accurate density / temperature variations, across multiple spectral channels or the temporal evolution of catastrophic cooling, leading to the chromospheric component of coronal rain. In this study, we reveal the multi-thermal and multi-stranded nature of flare-driven coronal rain at its source during its formation. We investigate the temporal evolution of the cooling curve of the post-flare arcade at the loop-top source, in order to better understand the catastrophic cooling process using temperature diagnostics from the X-Ray Sensor (XRS) onboard the Geostationary Operational Environmental Satellites (GOES), the Solar Dynamics Observatory / Atmospheric Imaging Assembly \citep[SDO/AIA:][]{2012SoPh..275...17L}, together with dynamics from H$\alpha$~656.3~nm \& Ca~{\sc ii}~854.2~nm spectral scans obtained, via  the CRisp Imaging Spectro-Polarimeter \citep[CRISP:][]{2008ApJ...689L..69S}, located at the Swedish 1-m Solar Telescope \citep[SST:][]{2003SPIE.4853..341S}.  

\par In section~2, we briefly outline the data reduction steps undertaken in this analysis of coordinated GOES, AIA and CRISP observations. In section~3, we present the results of our observations of flare-driven coronal rain and investigate the properties of the coronal rain source with CRISP. In section 4, we present our differential emission measure (DEM) analysis using multiple spectral channels in AIA to investigate the cooling processes (density / temperature variations both spatially and temporally) at the loop-top in the rain formation region, in the EUV. Finally in the discussion section, we investigate the radiative and conductive cooling timescales using a simplistic flare cooling model and discuss the implications of the DEM analysis, in the context of the combined observations of catastrophic cooling. 

\section{DATA REDUCTIONS}

\par We incorporate observations across a broad temperature range, using combined imaging and spectra at highest cadence and resolution (within the respective passbands), via the GOES (0.05-0.4~nm and 0.1-0.6~nm) soft X-ray light-curves, AIA (EUV) imaging in multiple spectral channels (see Fig.~\ref{fig2} for some of the spectral channels used in this study) and CRISP for imaging spectro-polarimetry in the visible / near-IR channels.

\par CRISP is a fast wavelength tuning, spectral imaging polarimeter that includes a dual Fabry-P{\' e}rot interferometer (FPI) and consists of a wideband and 2 narrowband cameras (transmitted and reflected), as described by \citet{2006A&A...447.1111S}. CRISP is especially suited for spectroscopic imaging of the chromosphere in the popular H$\alpha$~656.28~nm and Ca~{\sc ii}~854.2~nm spectral lines. CRISP is equipped with three high-speed, low-noise CCD cameras that operate at a frame rate of 36 fps. The spectral sampling is such that the transmission Full-Width-Half-Maximum (FWHM) of CRISP H$\alpha$ is 6.6~pm and the pre-filter is 0.49~nm. In these observations, the CRISP FOV (see Fig.~\ref{fig1}) was centred at [-349\arc,-329\arc] in solar-{\it x}/{\it y} on 1$^{st}$ July 2012 in the middle of AR~11515 and the observation sequence occurs between 15:08-16:31~UT. The CRISP wideband Field-Of-View (FOV) is co-aligned with SDO / AIA using the the background image (i.e. of the 170.0~nm continuum) of Fig.~\ref{fig1}, as a reference in the first time frame. The resulting FOV after clipping away CCD edge effects is 55\arc$\times$55\arc. The observation specifications consist of sequential spectral imaging at 6 wavelength points about the upper chromospheric H$\alpha$ 656.28~nm line centre (+/-0.1032~nm) followed by 9 wavelength points about the lower chromospheric Ca~{\sc ii}~854.2~nm line centre (+/- 0.0495~nm) followed by a full stokes sampling (Stokes I, Q, U and V) at 1 spectral position in the photospheric Fe~{\sc i}~630.2~nm line (-0.0048~nm) resulting in an effective cadence of 19~s (i.e. effectively a reduced cadence as a result of frame selection of the highest quality images). The image quality of the time series data significantly benefited from the correction of atmospheric distortions by the SST adaptive optics system \citep{2003SPIE.4853..370S}. Post-processing is applied to the data sets with the image restoration technique Multi-Object Multi-Frame Blind Deconvolution \citep[MOMFBD: ][]{2005SoPh..228..191V}. Consequently, every image is close to the theoretical diffraction limit for the SST with respect to the observed wavelengths. The pixel size of the H$\alpha$ images is 0.0597$\arcsec$. We followed the standard procedures in the reduction pipeline for CRISP data \citep{2015A&A...573A..40D}, which includes the post-MOMFBD correction for differential stretching as suggested by \citet{2012A&A...548A.114H}. We explore the fully processed datasets with CRISPEX (CRISP-EXplorer) \citep{2012ApJ...750...22V}, which is a versatile code for analysis of multi-dimensional spectral data cubes available through SolarSoft.


\begin{figure*}[!ht]
\centering
\includegraphics[clip=true,trim=0cm 1.5cm 0cm 1.5cm, scale=0.97, angle=0]{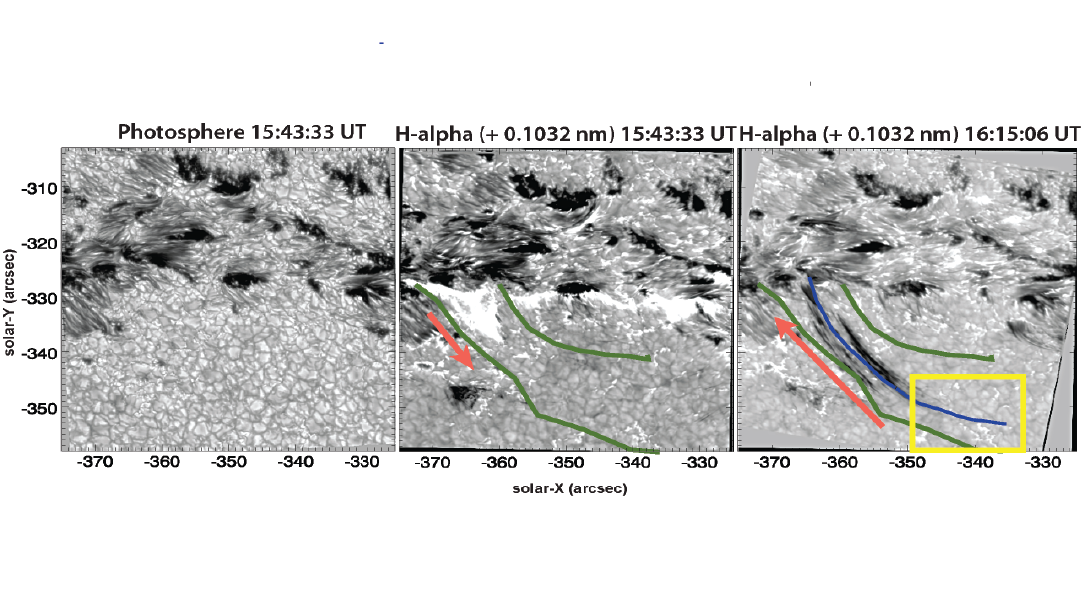}
\caption{We observe the H$\alpha$ spectral line at the far red-wing position (+0.1032~nm: middle \& right panels at different times) during ribbon formation at 15:43:33~UT (middle panel at solar-{\it y}: -330~arcsec) and later, of the chromospheric component of coronal rain at 16:15:06~UT (right panel). We present the photospheric Fe~{\sc i}~630.2~nm channel wideband image (left panel) at the time of chromospheric ribbon formation. We can spatially correlate the EUV post-flare loop boundary (as presented in Fig.~\ref{fig2}) overplotted as a green contour (middle and right panels). The blue curve marks the loop-leg cross-cut (right panel) used to produce the time-distance diagrams shown in Figs~\ref{fig7} \&~\ref{fig8}. The arrows depict the direction of propagation of the left-most ribbon structure during the ribbon formation (middle panel) and the subsequent direction of the coronal rain flow from the loop-top source (right panel), which is within the yellow box.}
\label{fig3}
\end{figure*}



\begin{figure*}[!ht]
\centering
\includegraphics[scale=1.20, angle=0]{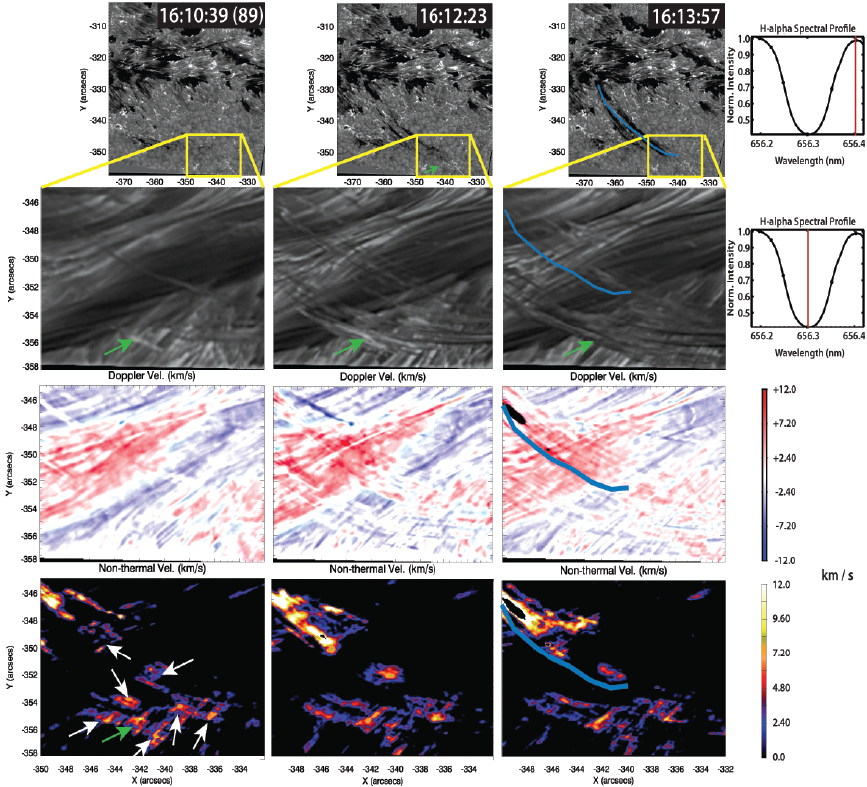}
\caption{{\it 1$^{st}$ row}: Within the CRISP full FOV (between 16:10:39~UT and 16:13:57~UT), we observe the formation of the dark coronal rain flow, along the trajectory of the post-flare loop arcade (as marked by the blue curve) within the grey-scale image of the H$\alpha$ far red-wing (+0.1032~nm). {\it 2$^{nd}$ row}: We present the zoomed FOV, as outlined by the yellow box, for the H$\alpha$ line core. The blue curve is again overlaid in the 3$^{rd}$ panel for perspective and we reveal the evolution of the first detected loop-top coronal rain source with the green arrows. The spectral positions denoting the grey-scale images of the 1$^{st}$ and 2$^{nd}$ rows, are marked as red lines within the H$\alpha$ spectral scan in the 4$^{th}$ panels of these rows. {\it 3$^{rd}$ row}: Dopplergrams are presented, as derived from the H$\alpha$ spectral profile for each spatial pixel, from within the yellow-box FOV. The spectral resolution is 6~pm providing an error in the velocities of ~$\pm$0.3\kms. {\it 4$^{th}$ row}: The background subtracted (i.e. using before and after images of the coronal rain formation region to remove the presence of underlying fibrils) line-width broadening maps are used to derive the non-thermal speed from measuring the FWHM change between the H$\alpha$ spectral profiles at the location of the rain formation and the quiet sun reference profile. The first detected rain source marked with a green arrow.}
\label{fig4}
\end{figure*}


\par The GOES X-ray observations are obtained using the GOES graphical user interface, within the SolarSoft  ({\small SSWIDL}) routines and processed for cleaning and background subtraction\footnote{We followed the TEBBS procedure for Temperature and Emission measure-Based Background Subtraction, outlined here: http://www.solarmonitor.org/tebbs/about/} \citep{2012ApJS..202...11R} of soft X-ray flux, temperature and emission measure (EM) light-curves (as presented in Figs.~\ref{fig2}~\&~\ref{fig18} for the 0.05-0.4~nm passband). Similarly, AIA data are processed from level 1.0 to level 1.5 using standard {\small SSWIDL} routines to corrected for dark and flat fielding, plate-scale corrections and limb fitting for alignment between the AIA channels with 170.0~nm (a reference channel for co-alignment). SDO/AIA observes with a cadence of 12~s and a pixel size of 0.6\arc\ (corresponding to a spatial resolution of $\sim$1.6\arc\ ). With the level 1.5 data product it is expected that there is a small spatial offset between the internal AIA channels on the order of 0.25 to 0.5 AIA-pixels \citep{2011ApJ...732...81A}. To achieve sub-AIA pixel accuracy in the temporal and spatial co-alignment of CRISP images with AIA, we cross-correlate the clearly identifiable active region bright points. The AIA images are then derotated to that time frame and the bright points are tracked in time with CRISP enabling the excellent co-alignment throughout the observation. 

    
\begin{figure*}[!ht]
\centering
\includegraphics[scale=0.9, angle=0]{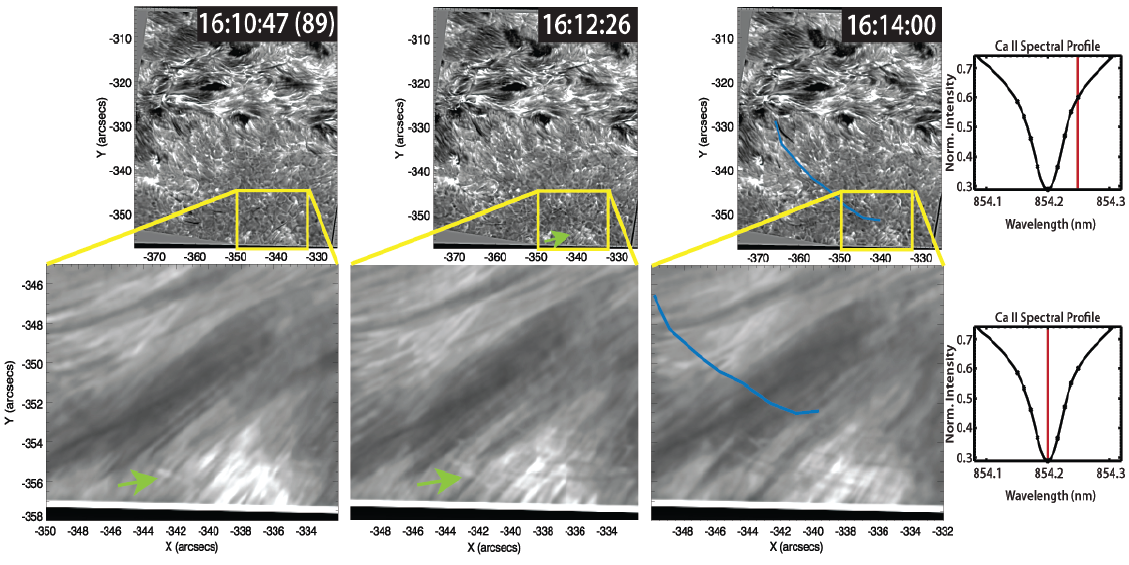}
\caption{{\it Top row}: The grey-scale images, as observed in the Ca~{\sc ii}~854.2~nm red-wing (+0.0495~nm) channel, are presented for comparison with Fig.~\ref{fig4}. We mark the blue cross-cut curve for reference with the H$\alpha$ images of Fig.~\ref{fig4} and the yellow box of the coronal rain formation region. {\it Bottom row}: As compared with Fig.~\ref{fig4}, we present the Ca~{\sc ii}~854.2~nm line core images and the green arrows indicate the coronal rain source signature in emission. The Ca~{\sc ii}~854.2~nm spectral profile is presented in the 4$^{th}$ panels in each row with the corresponding spectral positions for the images, marked with a red line.}
\label{fig5}
\end{figure*}



\begin{figure}
\centering
\includegraphics[scale=0.16, angle=0]{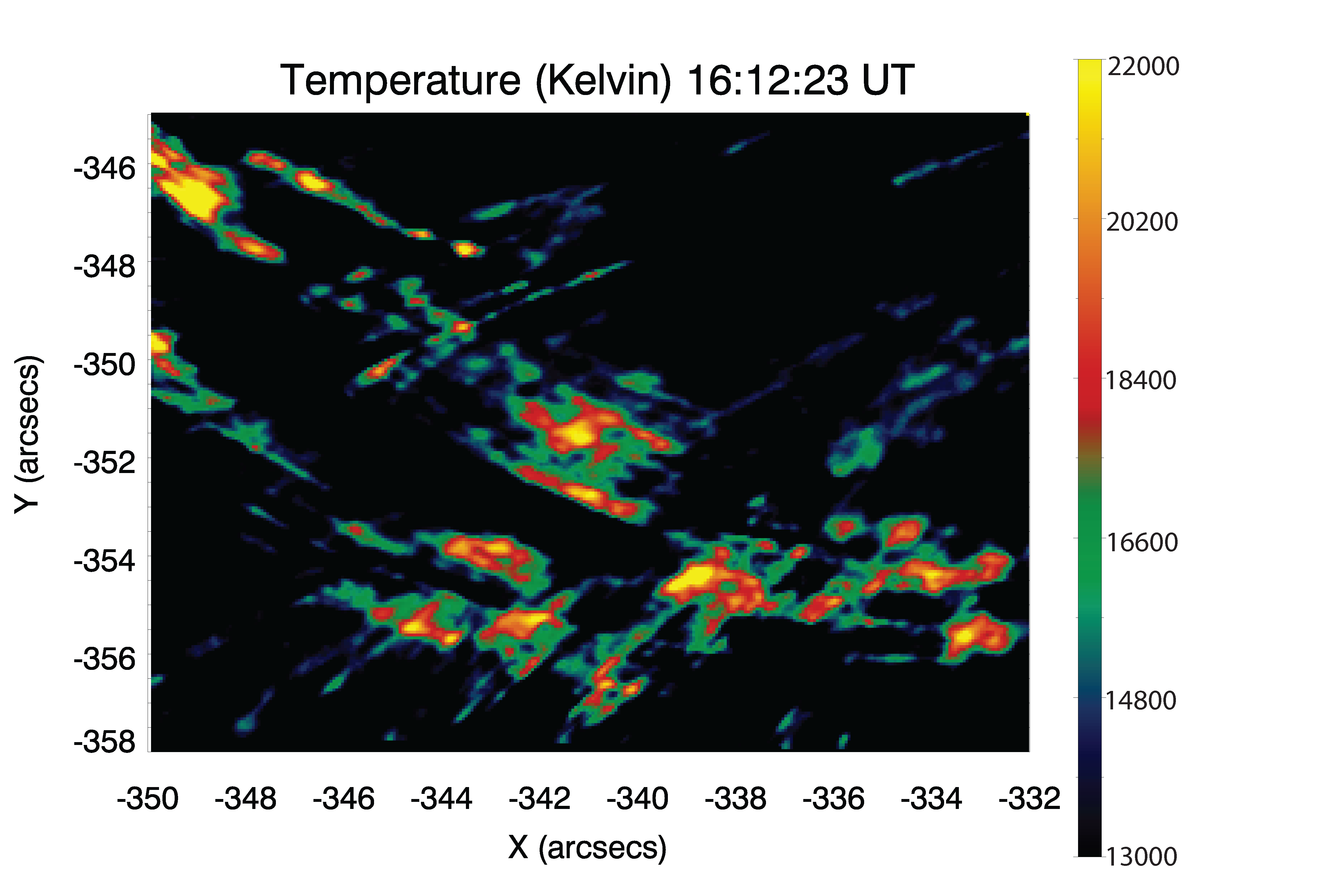}
\caption{The temperature map of the yellow boxed region (i.e. defining the H$\alpha$ coronal rain formation region outlined in previous figures) is presented for 16:12:23~UT (i.e. 124~s after the first appearance of a source of coronal rain). Multiple (at least 8) relatively hot sources ($\ge$22000 Kelvin) of coronal rain near the loop-top. These temperatures represent an upper limit of the true temperature, obtained assuming negligible non-thermal velocity. A background subtraction has been applied to remove the presence of underlying fibrils using before and after images of the rain formation (NB The background fibrils are longer-lived features, hence, easily subtracted).}
\label{fig6}
\end{figure}



\begin{figure}[!h]
\includegraphics[clip=true,trim=0cm 0cm 0cm 0cm, scale=0.17, angle=0]{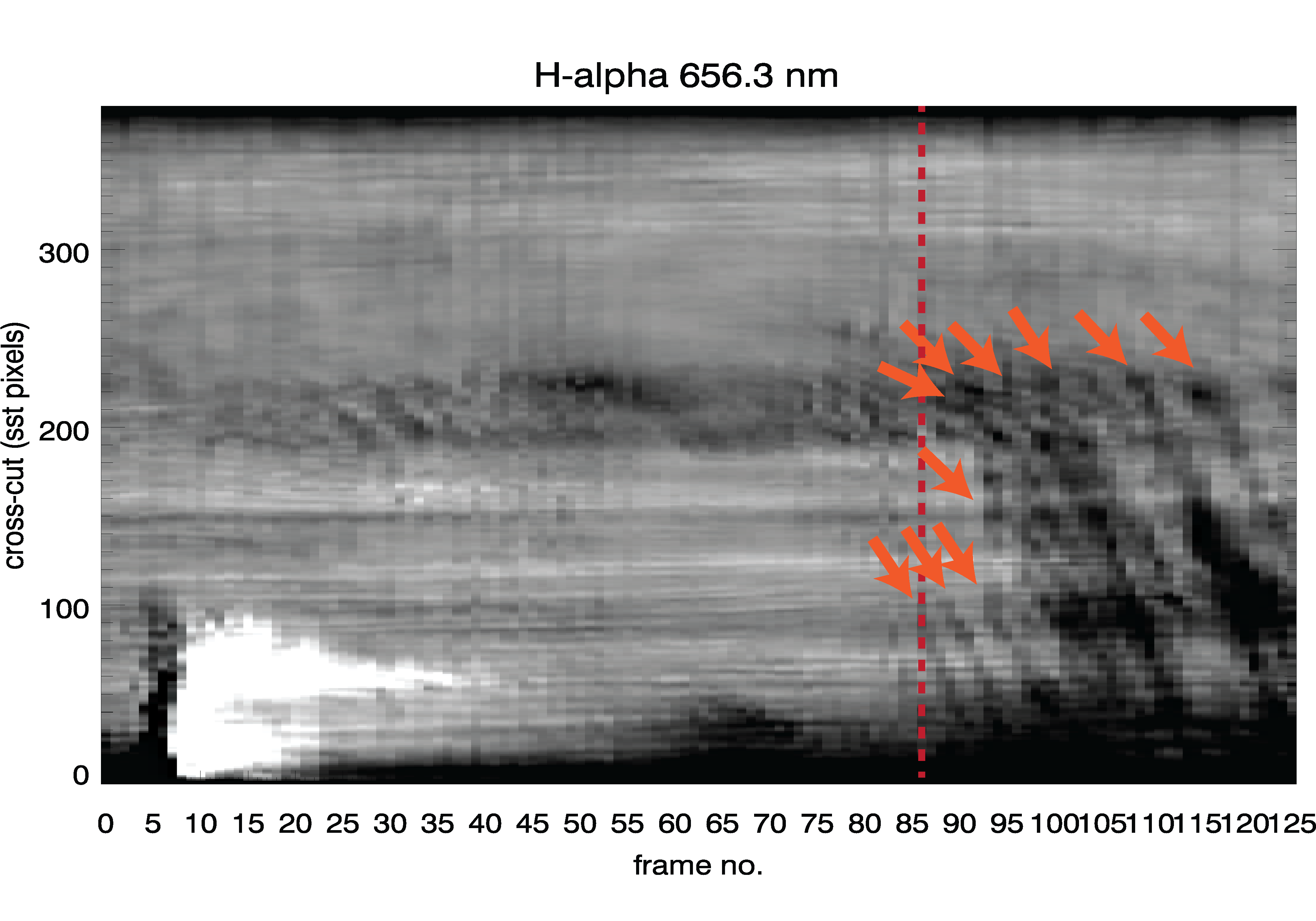}
\caption{Time-distance diagram from the H$\alpha$ far red-wing (+0.1032~nm) FOV, using data extracted along the blue curve and averaged over multiple locations as the blue curve was scanned between the boundary of the EUV post-flare loop, as bounded by the green contours in Fig.~\ref{fig3}. The diagram therefore reveals the changes in the chromospheric rain flows on average across the width of the EUV loop. The {\it y}-axis (distance along the curve from foot-point to loop-top) increases from zero (i.e. the apparent foot-point in the flare ribbon) to 370 SST-pixels, i.e. the apparent loop-top, where the coronal rain formation is located within the yellow boxed region. The {\it x}-axis (time) increases from a frame corresponding to 15:41:20~UT (i.e. during the pre-flare phase). The red arrows mark the formations of individual rain strands along the loop leg after loop-top rain formation. The vertical red dashed line is the time frame corresponding to the first signature of the loop-top source (as presented in Fig.~\ref{fig10a}.}
\label{fig7}
\end{figure}


\section{OBSERVATIONS OF FLARE-DRIVEN CORONAL RAIN}

\par The C8.2-class solar flare impulsive phase lasted for 3~min 10~s according to the GOES light-curve in the 0.05-0.4~nm channel for soft X-ray emission, as presented in Fig.~\ref{fig2} ({\it Main}). The post-flare decay phase in X-ray emission lasted for 18~mins, after it reached a peak at 15:44:40~UT, then the EUV post-flare arcade became visible in the hottest AIA channels, as is presented in Fig.~\ref{fig2} ({\it inset}). During the formation of EUV post-flare loops (i.e. between 16:05~UT and16:23~UT), we detected a strong flow of multiple coronal rain strands in the chromospheric H$\alpha$ line, which appears to fall back towards the surface within the EUV loop cross-sections. We identify the origins of this flow to lie within the yellow boxed region of Figs.~\ref{fig2}~\&~\ref{fig3}. This region is the focus of further investigation in this study into the nature of coronal rain formation. The multiple, dark coronal rain strands are presented in Fig.~\ref{fig3} (right panel). The left panel shows the wideband image from the Fe~{\sc i}~630.2~nm spectral window which represents the signal at the photospheric surface. It is clear that there is no white-light signal associated with the flare which is common for weaker (C-class) flares. In Fig.~\ref{fig3}, we clearly demonstrate that the source and sink of flare-driven chromospheric plasma in coronal rain, i.e. ultimately the origins of its existence, lies in the bright H$\alpha$ ribbon formation (foot-point heating), visible within the red-wing images of H$\alpha$ (middle panel). The time delay between the flare ribbon formation and the first appearance of coronal rain formation (frame no. 88, 16:10:19~UT) is $\sim$26~mins and by 16:15:06~UT we detect a full development of the rain strands further along the loop-leg, returning to the location of the ribbons. In order to investigate the properties of the formation of the return flow as coronal rain from the loop-top we use the H$\alpha$ line core imaging which reveal the structural details of the chromospheric plasma. 

\par In Fig.~\ref{fig4}, we zoom into the yellow boxed region of the coronal rain source and present the H$\alpha$ line core images ({\it 1$^{st}$ row}) for frame no. 89 (16:10:39~UT). This loop-top source is initially in emission in H$\alpha$ and located with the green arrow. At 16:12:23~UT this bright source cools into an absorption profile and becomes more extended spatially (to the left and right of the loop-top) by 16:13:57~UT, as it proceeds to fall back towards the foot-points. This bright source is very localised and initially has a circular cross-section of $\sim$11000~km$^{2}$ before becoming more elongated by 16:12:23~UT. By 16:13:57~UT, we can detect at least 8 parallel, darker (cooler) strands at the loop-top. When we consider the Doppler velocity of the H$\alpha$ flow field at the source of the coronal rain ({\it 3$^{rd}$ row}), we do not detect any net flows at the loop-top sources. Only along the path of the blue curve do we detect a strong net red-shift of 12.0~$\pm$0.3~km~s$^{-1}$ which corresponds to the location of the darkest rain strands in the far red-wing of H$\alpha$ ({\it 1$^{st}$ row} 3$^{rd}$ panel). The thermal velocity maps ({\it 4$^{th}$ row}) are determined from the Doppler broadening of the spectral scans per pixel. The method used to determine these velocities involved subtracting the background through using image frames from before and after the formation of the rain within the yellow box region. Then we determined a reference profile through averaging the spectral profiles per pixel in a small region away from the rain formation where there is no activity within the interval of the rain formation (i.e. from 16:05~UT-16:23~UT). We then measured the Doppler shifts and FWHM per pixel in the rain strands relative to the reference profile. The alternative method of fitting a Gaussian profile to the spectral profile in H$\alpha$ is not so effective given the relatively low number of spectral positions scanned. There is a strong thermal broadening in the range of 4-12~$\pm$0.3~km~s$^{-1}$, at the location of the bright H$\alpha$ loop-top sources of coronal rain ({\it 4$^{th}$ row}). We can estimate an upper limit for the corresponding gas temperature of the emitting region from the line width broadening in the rain formation with

\begin{equation}
\mathord{T} = 
\frac{1}{8\,\,\text{ln}2}\frac{c^{2}m_{H}}{k_{B}}\left(\frac{\text{FWHM}}{\lambda_{0}}\right)^{2},
\label{halpha_temp}
\end{equation}


\par where $k_{b}$ is the Boltzmann constant, $\lambda_{0}$ is the rest wavelength of the reference profile, $m_{H}$ is the mass of a Hydrogen atom and here we ignore micro turbulence physics. The resulting temperature map for the yellow boxed region of Fig.~\ref{fig4} is presented in Fig.~\ref{fig6}. To determine the non-thermal speed from Fig.~\ref{fig4} we used a rearrangement of Equation~\ref{halpha_temp} and we assume that the Doppler width of the reference profile represents the quiet Sun formation temperature of H$\alpha$ line of $1\,\times\,10^{4}$~K. 

\par In Fig.~\ref{fig5}, in a similar fashion as with Fig.~\ref{fig4}, we investigate the lower chromospheric Ca~{\sc ii} 854.2~nm line core signatures of the loop-top rain formation within the yellow boxed region. We do not detect any signatures of red-shifted dark coronal rain strands in Ca~{\sc ii} 854.2~nm at the corresponding times. However, we detect a bright coronal rain source, marked with the green arrows, which are on average smaller in cross-section and relatively less bright (with respect to its background) compared with the same H$\alpha$ bright source (compared with its background level) for the same time frame. In the line core images we detect the faint signatures of absorption of rain strands at the loop-top, extending away from the bright source, in the same locations as the dark strands from H$\alpha$. This implies that the temperature and density properties of coronal rain at the loop-top may be much more structured at the loop-top than along the loop leg. The detection of Ca~{\sc ii} coronal rain source features implies that the plasma temperature must cool further, below H$\alpha$ formation temperatures, in agreement with measurements of quiescent coronal rain in \citep{2015ApJ...806...81A}. We do not detect anything significant within the Dopplergrams or line width broadening maps for Ca~{\sc ii} because the signal in the rain is so weak on-disk.  

\par
\begin{figure}[!h]
\centering
\includegraphics[clip=true,trim=7cm 0cm 0cm 0cm, scale=0.34, angle=0]{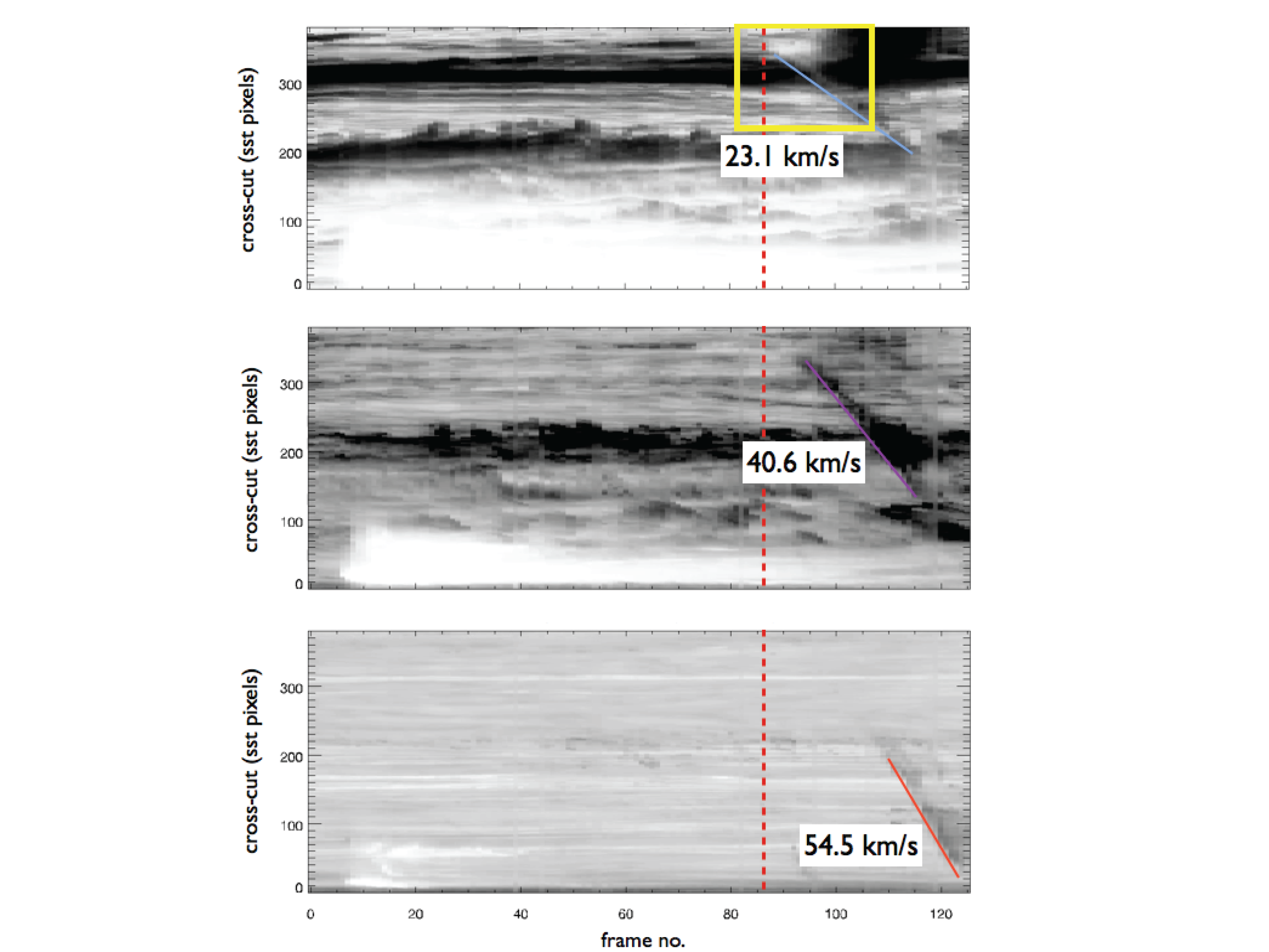}
\caption{Time-distance diagrams from the H$\alpha$ line core ({\it top}), near red-wing (+0.0774~nm: {\it middle}) and far red-wing (+0.1032~nm: {\it bottom}) spectral images, using the data along the blue curve from Fig.~\ref{fig3} (right panel). The vertical red dashed line is the time frame corresponding to the first signature of the loop-top source (as presented in Fig.~\ref{fig10a}). The loop-top coronal rain bright emitting source (topside within the yellow box) and the subsequent formation of the rain strands in absorption (i.e. blue to purple to red solid lines) become more red shifted in time and the acceleration decreases in time along the trajectory of the EUV loop-leg.}
\label{fig8}
\end{figure}

 From Fig.~\ref{fig6}, we map the temperature distribution and find that the hottest components of the source of the rain, corresponding to the brightest structures, is at most 20000-22000~K. When we consider plasma at 14000-16000~K we can detect the faint outline of the post-flare loop-arcade in the chromospheric plasma which may indicate that the chromospheric plasma remains thermally confined to the magnetic structure of the loop system in partial ionisation. As mentioned previously, the flow field in the H$\alpha$ coronal rain is traced out with a blue curve in Fig.~\ref{fig3} ({\it right panel}). From this blue curve, we extract time-distance plots to learn more about the nature of the coronal rain flow field, as presented in Figs.~\ref{fig7}~\&~\ref{fig8}. The one-to-one correspondence of these sources with the EUV loop activity will be discussed later in this section.

\par In Fig.~\ref{fig7}, we measure the flow field properties of the coronal rain strands (beyond frame no.~86, marked by the vertical red dashed line: 16:09:44~UT), as they fall back towards the post-flare foot-point within the H$\alpha$ far red-wing (+0.1032~nm) FOV.  The loop half-length ($L$) is 32~$\pm$0.4~Mm and is traced by the blue-curve cross-cut used to extract the time-distance plots. NB The loop half-length was determined through measuring the separation of the foot-points of the post-flare arcade, using the mid-point of the loop cross-section as observed in the 17.1~nm channel from AIA, at the time of coronal rain formation in H$\alpha$ (i.e. see the contoured loop arcade in 17.1~nm in the inset panel of Fig.~\ref{fig2}). This foot-point separation is measured as 56~arcsec. We then determined the loop half-length from a half circle assuming a circular post-flare loop has formed between the foot-points. At frame no.~88, we detect the onset of red-shifted, dark flows along the loop-leg which flow back to the foot-point. The bright flare ribbon exists at the foot-point until frame no.~40 and extends outwards along the loop by $\sim$4300~km, implying an outward propagation of this heating signature into the post-flare loop system. The longest continuous coronal rain strands detected here are $\sim$10,700~km, assuming we do not need to consider projection effects in this estimation. The rain strands appear to fall back within a range of velocities spanning 52-64~km~s$^{-1}$ and exhibit periodicity. We can detect 10 sequential strands (marked with arrows) which appear to last for typically 3-4 time frames (corresponding to 55-70~s) and are separated by a similar time interval. The rain shower (a termed first coined by \citet{2012ApJ...745..152A} describing a sudden onset of multiple rain strand formations) appears to end at frame no.~125 (16:22:38~UT) resulting in a shower lifetime in the range of 770-780~s.


\begin{figure*}[!ht]
\includegraphics[clip=true,trim=0cm 0cm 0cm 0cm, scale=0.36, angle=0]{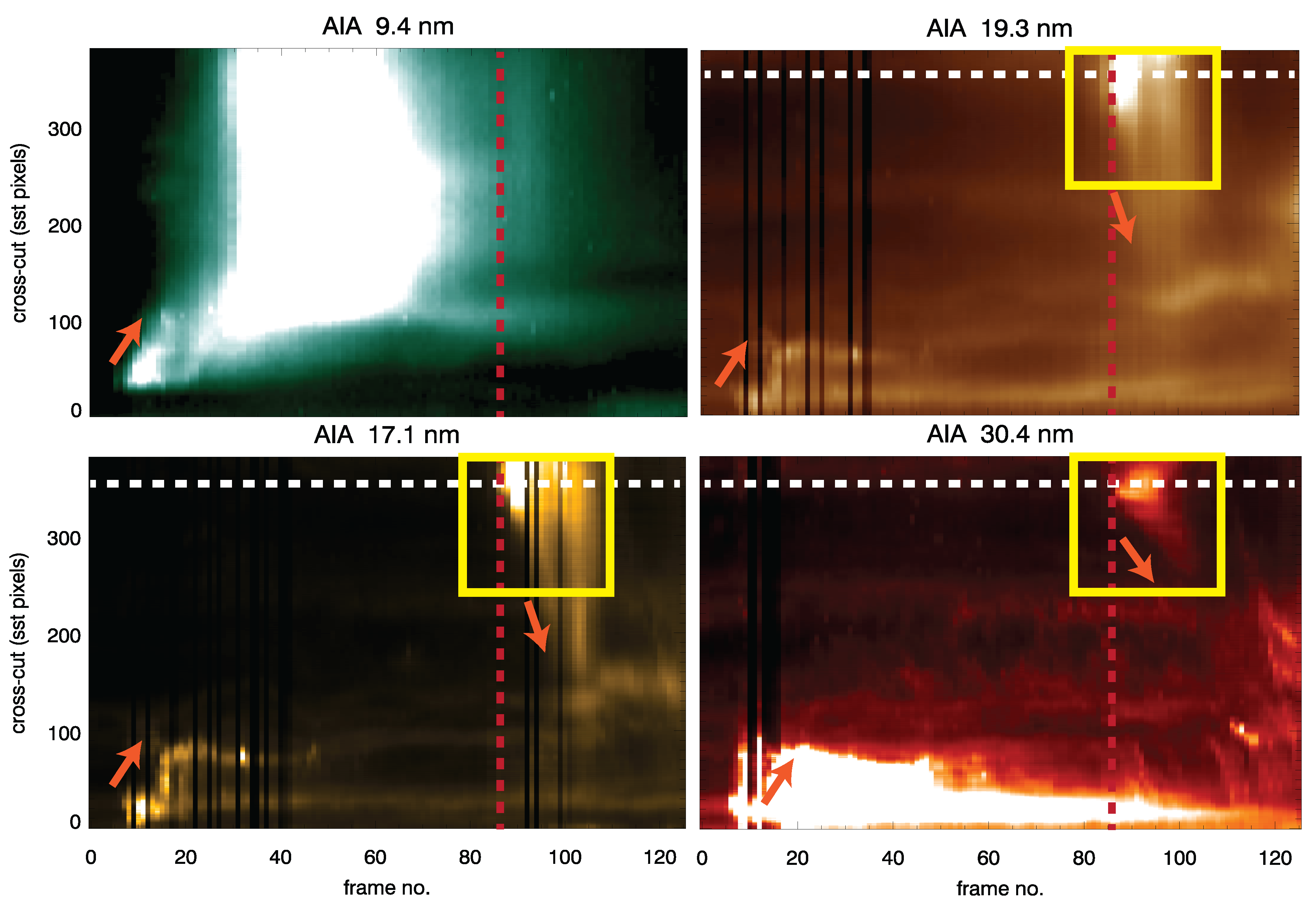}
\caption{Time-distance diagrams are extracted along the same blue curve (as a direct comparison with Fig.~\ref{fig8}) but for co-spatial and co-temporal AIA flaring spectral channel 9.4~nm ({\it top left}), coronal channels 19.3~nm ({\it top right}) and 17.1~nm ({\it bottom left}), as well as, the transition region (TR) channel 30.4~nm ({\it bottom right}). We detect the bright coronal rain formation source (bounded by the yellow box) at the loop-top and the subsequent rain strands / flows down the loop using downward-pointing orange arrows. The upward-pointing orange arrows represent the outward propagation of emission in the flare ribbon source in AIA along the loop early on. The red vertical dashed line represents frame no. 86, which is the time we first detect the hottest coronal rain source. The white horizontal dashed line, which passes through the coronal rain source, is used to produce the light curve of Fig.~\ref{fig10}, for all channels under discussion.}
\label{fig9}
\end{figure*}



\begin{figure}[!h]
\includegraphics[clip=true,trim=0cm 0cm 0cm 0cm, scale=0.175, angle=0]{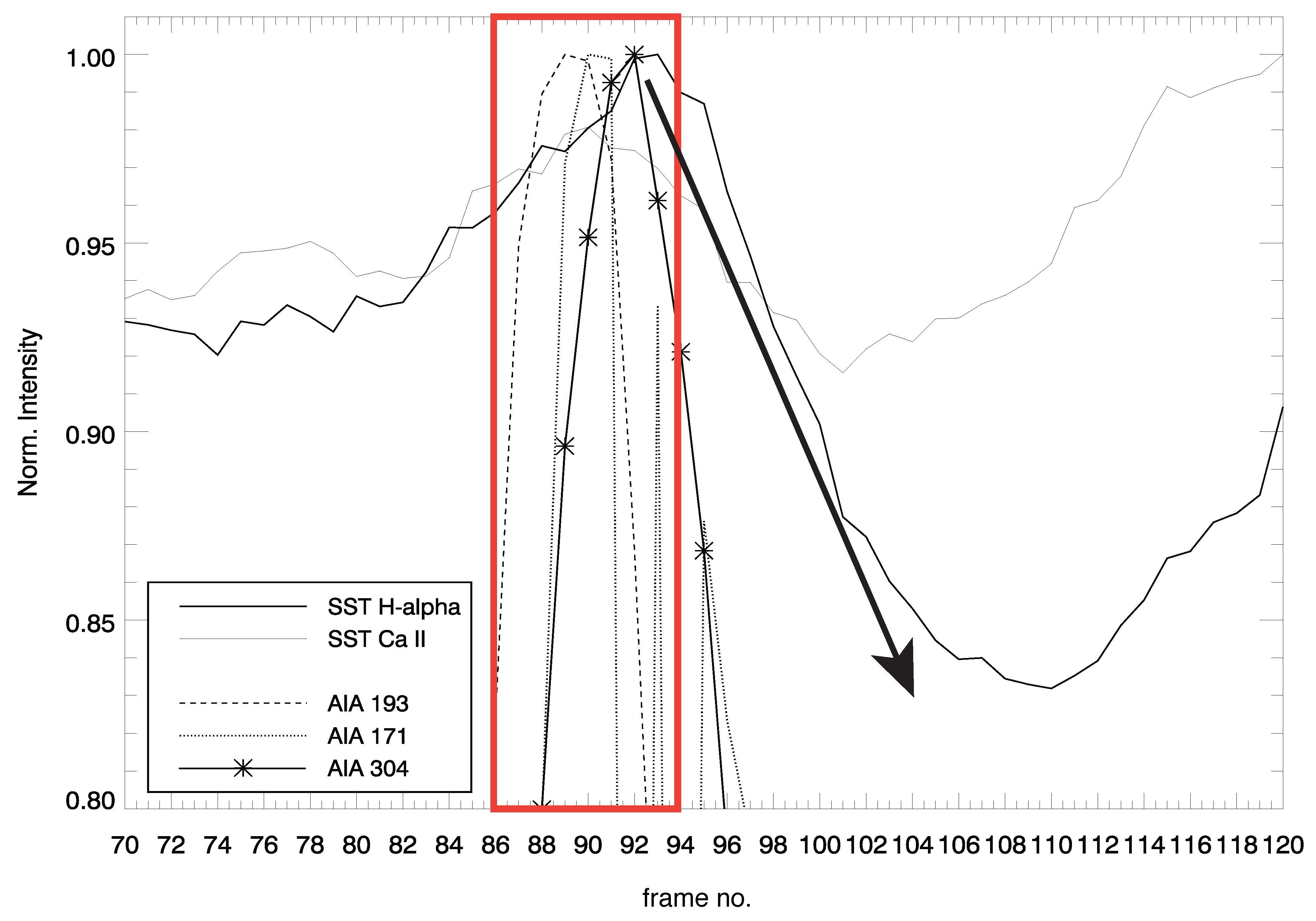}
\caption{Light-curves are extracted from the time-distance diagrams of Fig.~\ref{fig9} (see horizontal white-dashed line) for AIA channels 19.3~nm, 17.1~nm, 30.4~nm and overlaid with CRISP H$\alpha$ 656.28~nm and Ca~{\sc ii}~854.2~nm channels from Fig.~\ref{fig8}. The legend distinguishes the lines corresponding to different channels. The black arrow marks the onset of the chromospheric component of coronal rain and resulting decrease in the intensity at the loop-top source, in both H$\alpha$ and Ca~{\sc ii}.}
\label{fig10}
\end{figure}


\par In Fig.~\ref{fig8}, we consider three time-distance diagrams for the blue curve track derived from images different spectral positions in H$\alpha$. We detect the initial, bright coronal rain source that was also present in the FOV from Fig.~\ref{fig4}. At +0.1032~nm the rain strand exists closer to the loop foot-point (50-200 on the {\it y}-axis) whereas at +0.0774~nm the same strand is detected at an earlier time and closer to the loop-top (120-300 on the {\it y}-axis). Furthermore, the loop-top component of the rain can be detected in H$\alpha$ core between 200-360 and appears to originate as a bright source (within the yellow box) at frame no.~88. The coronal rain flow extending from this bright source (highlighted with the blue solid line) has an apparent velocity of 23.1~$\pm$2~km~s$^{-1}$. In the subsequent panels, scanning further into the red-wing, we detect progressively faster flows (i.e. highlighted with purple and red solid lines) corresponding to 40.6~$\pm$2~km~s$^{-1}$ and 54.5~$\pm$2~km~s$^{-1}$, respectively. The coronal rain flow overall appears to accelerate in time, i.e. between frame no. 88 (red vertical dashed line) at the loop-top until it reaches the foot-point at frame no.~125. However the rate of change of acceleration decreases from the loop-top to half-way along the loop-leg at 76~m/s$^{2}$, to 60~m/s$^{2}$ from the loop-leg to the foot-point. As the coronal rain clumps fall along the loop arcade, they encounter an increasingly more dense atmosphere at the loop foot-points in the TR and on into the chromosphere. The reduced acceleration in the rain has also been observed in quiescent coronal rain \citep{2012ApJ...745..152A} and this was confirmed numerically to be due to the increase of gas pressure in the lower atmosphere with the greater local densities \citep{2013ApJ...771L..29F}.


\begin{figure*}[!ht]
\includegraphics[clip=true,trim=0cm 0cm 0cm 0cm, scale=0.37, angle=0]{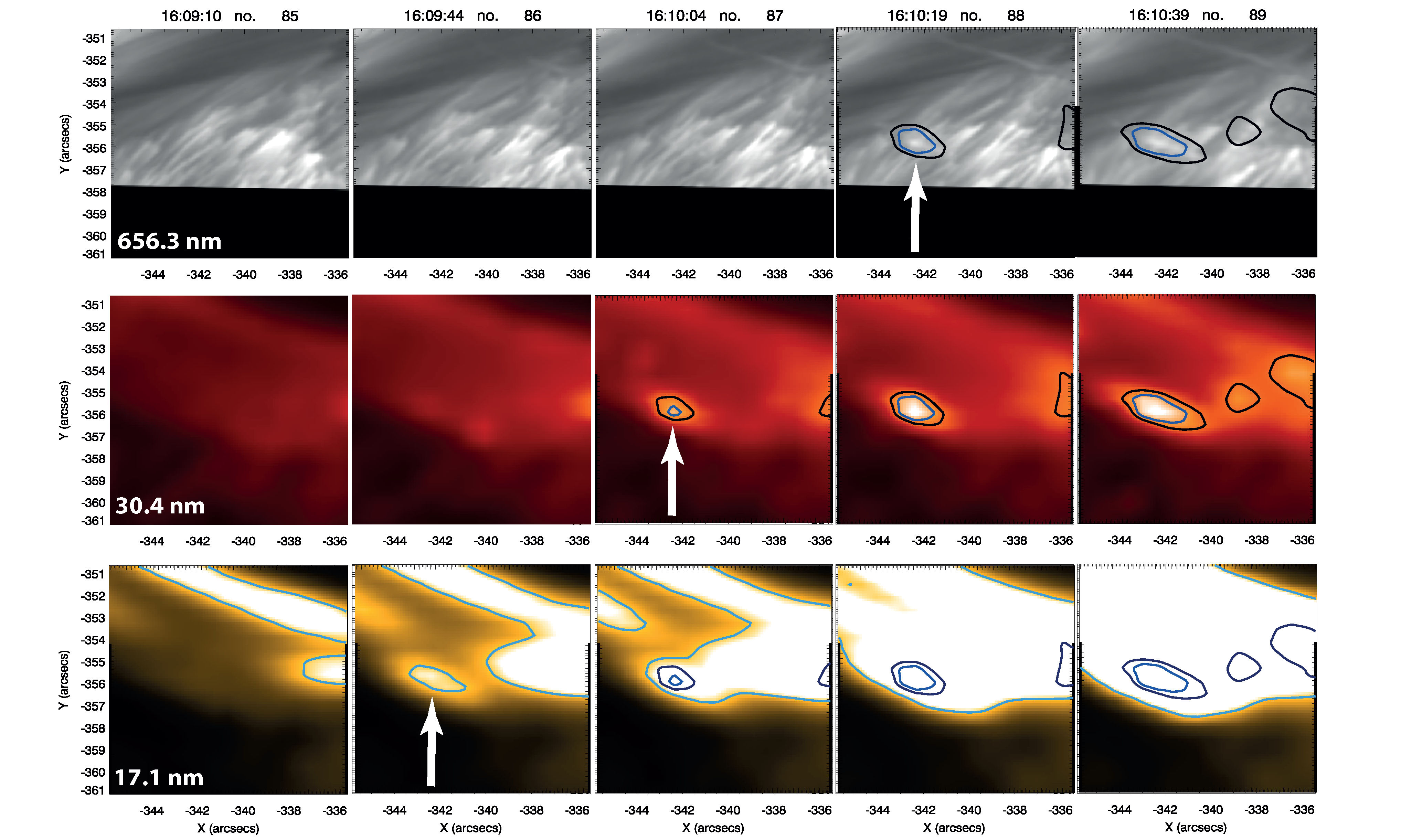}
\caption{We reveal the multi-thermal, co-spatial signatures of the coronal rain loop-top source at the time of its formation. The white arrows mark the location of the rain source in each spectral channel when it first appears, respectively. The arrows demonstrate the order of appearance of the source, becoming detectable in the lower temperature channels. The contours of the 30.4~nm ({\it middle row}) channel are overlaid onto the H$\alpha$ core images ({\it top row}) and, similarly, they are superimposed together with 17.1~nm contours on the 17.1~nm images ({\it bottom row}). The temporal evolution of the multi-thermal properties of this loop-top rain source, corresponding to the peak temperature of the respective passbands, appear as star symbols within the summarising light-curve of Fig.~\ref{fig18}.}
\label{fig10a}
\end{figure*}


\par In Fig.~\ref{fig9}, we can detect the ribbon formation in the early time frames in all AIA EUV channels (i.e. using the same cross-cut curve). This is most notable in the hottest AIA channel 9.4~nm, where we detect the continuation of the coronal flaring plasma, along the length of the curve towards the loop-top, between frame no.~65 and frame no. 86. When we look at the equivalent time-distance diagrams from the EUV channels for 19.3~nm, 17.1~nm and 30.4~nm images, we can identify the same bright coronal rain source in the EUV within the yellow box at the loop-top. From this source, again we detect bright flows extending away (marked with red arrows) and returning to the foot-point, in conjunction with the H$\alpha$ strands in the line core time-distance diagram. In particular we detect more clearly distinct bright tracks in the TR channel of 30.4~nm further along the loop-leg. The hottest signatures at the coronal rain source appear first at frame no.~86 (as marked with the red dashed line), i.e. 2 frames prior to the first detection in the cooler chromospheric lines. The co-location of the hot and cool components of the sources needs to be carefully considered due to line-of-sight effects. A better estimation of the temperature evolution from a DEM analysis, incorporating the intensity contributions from all AIA flaring channels, will be carried out in the next section.


\begin{figure}[!ht]
\includegraphics[clip=true,trim=0cm 1.5cm 0cm 2cm, scale=0.35, angle=0]{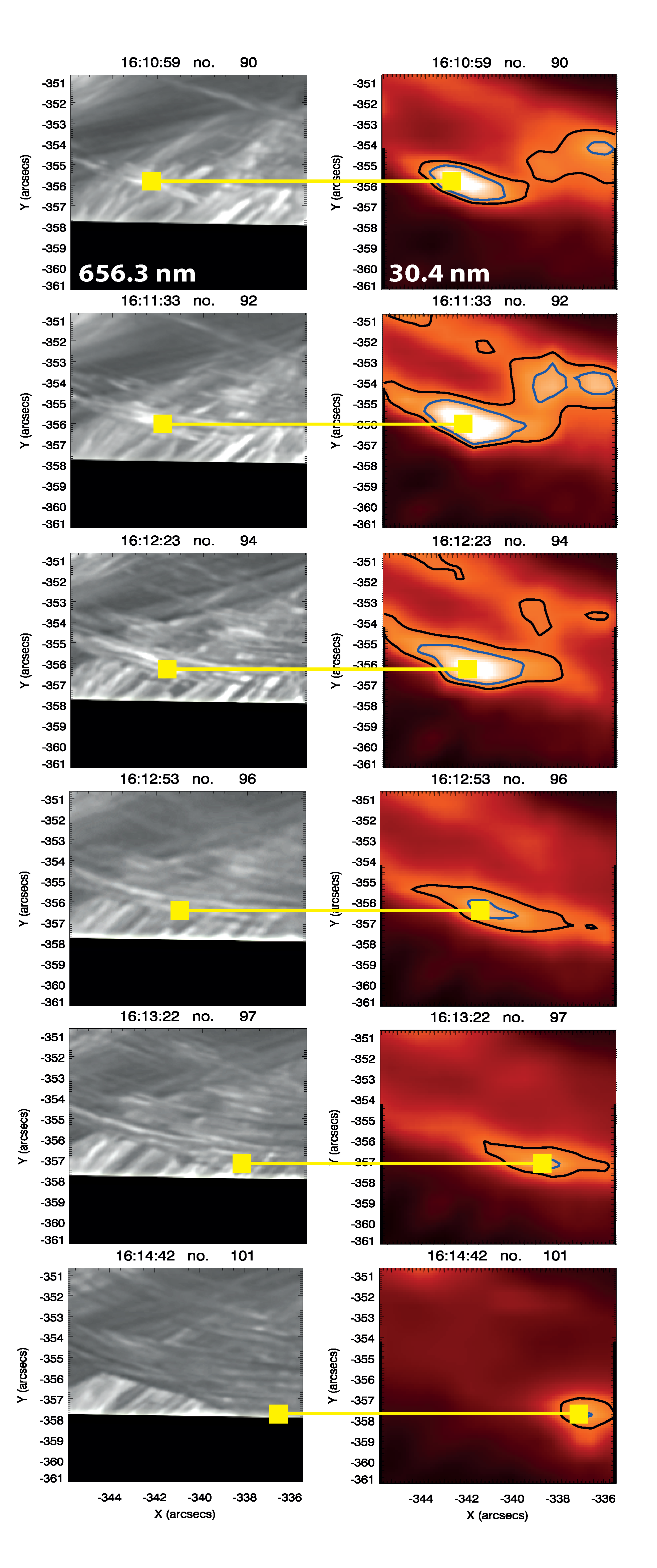}
\caption{We present the co-spatial, multi-thermal flow of a bright H$\alpha$ loop-top coronal rain source after its formation (as presented in Fig.~\ref{fig10a}). The images evolve in time from top (16:10:59~UT) to bottom (16:14:42~UT) and we present the H$\alpha$ core images in grey-scale on the left-side and 30.4~nm images on the right-side. The corresponding flows in both channels, along the loop from left to right (spatially in the FOV), is represented using the horizontal yellow solid lines in each row.}
\label{fig10b}
\end{figure}


\par The white horizontal dashed line in Fig.~\ref{fig9} is extracted to produce the light-curves of Fig.~\ref{fig10}. In Fig.~\ref{fig10}, we present the normalised intensity light-curves of the EUV and chromospheric signatures for comparison, during the rain formation. The separation of the peaks of the respective channels in the EUV highlights the previous point regarding the presence of a rapid cooling process, only a few minutes. Between time frame no. 86 and 93 (bounded by the red box) we observe the peak in the EUV, together with the first appearance and increasing brightening of the coronal rain source in H$\alpha$ and Ca~{\sc ii} before absorption and decreased intensity after frame no. 93. The separation of the bright EUV peaks during the formation of the coronal rain at the loop-top is on average 27~s.


\begin{figure*}[!ht]
\centering
\includegraphics[clip=true,trim=0cm 0.5cm 0cm 0cm, scale=1.76, angle=0]{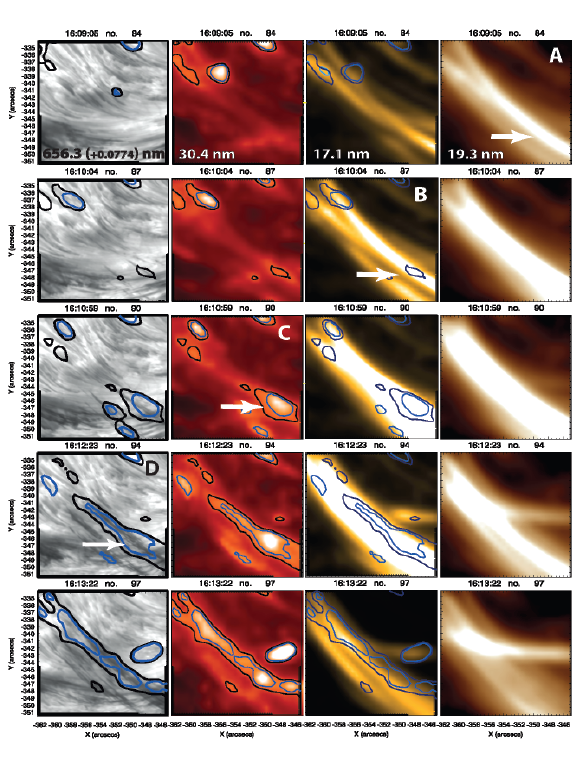}
\caption{We present the co-spatial, multi-thermal flow of another coronal rain strand from a bright H$\alpha$ loop-top coronal rain source, along an opposing loop-leg (i.e. compared with the loop-leg strand from Fig.~\ref{fig10b}). Temporal evolution is from top-to-bottom (i.e. from 16:09:05~UT to 16:13:22~UT) and the co-spatial spectral images evolve from the visible (H$\alpha$ red wing image at +0.0774~nm) into the hotter EUV channels (30.4~nm, 17.1~nm and 19.3~nm) from left-to-right panels. The white arrows depict the first clear detection of the rain source within the loop-leg, in each respective channel. The trajectory of the loop-leg is very bright and well-defined within the 19.3~nm image at 16:09:05~UT. This evolution though the spectral channels is also labelled {\bf A-D}, in the respective panels. The contours of 30.4~nm are overlaid in all panels (except 19.3~nm panels) for context. Time stamps concerning the formation of this coronal rain source and strand, in each spectral channel, appears as symbols within the summarising light-curve of Fig.~\ref{fig18}.}
\label{fig10c}
\end{figure*}


\par In Fig.~\ref{fig10a}, we present the spatial correspondence between the discretised location of the initial H$\alpha$ rain bright loop-top source (contoured using 30.4~nm intensity contours) at 16:10:19~UT (frame no. 88) and its hotter multi-thermal components in 30.4~nm and 171.~nm, which appear prior to the chromospheric component at 16:10:04~UT in 30.4~nm and 16:09:44~UT in 17.1~nm. From imaging in narrow band spectral channels there is clearly evidence for cooling of the rain source from the EUV to the visible channels, which appears within the contoured region and the formation region is highlighted with the white arrows. The subsequent increase in the spatial size of the source and its intensity in all channels suggests we can expect a corresponding increase in the EM at this source, as gradually more plasma condenses to the coronal / TR temperatures. The rain source appears as a spatially discretised source in the coronal channel 17.1~nm at 16:09:44~UT and this relatively faint (contoured) source becomes more intense, more elongated along the trajectory of the loop then multiple sources become detectable by 16:10:39~UT.  A similar growth rate of the source (spatially) occurs in the 30.4~nm channel and later in the H$\alpha$ channel. This one-to-one correspondence between the spectral channels, its co-spatialiity and the corresponding spatial evolution, leads us to suggest that this source in the visible channels is indeed co-located (in 3D space) with the EUV channels and, therefore, we demonstrate the multi-thermal nature of coronal rain at the source of its formation. In other words, the given rain strands must be structured with both hot and faint, as well as, cool and dense plasma. 


\begin{figure*}[!ht]
\includegraphics[clip=true,trim=0cm 4cm 0cm 3cm, scale=0.49, angle=0]{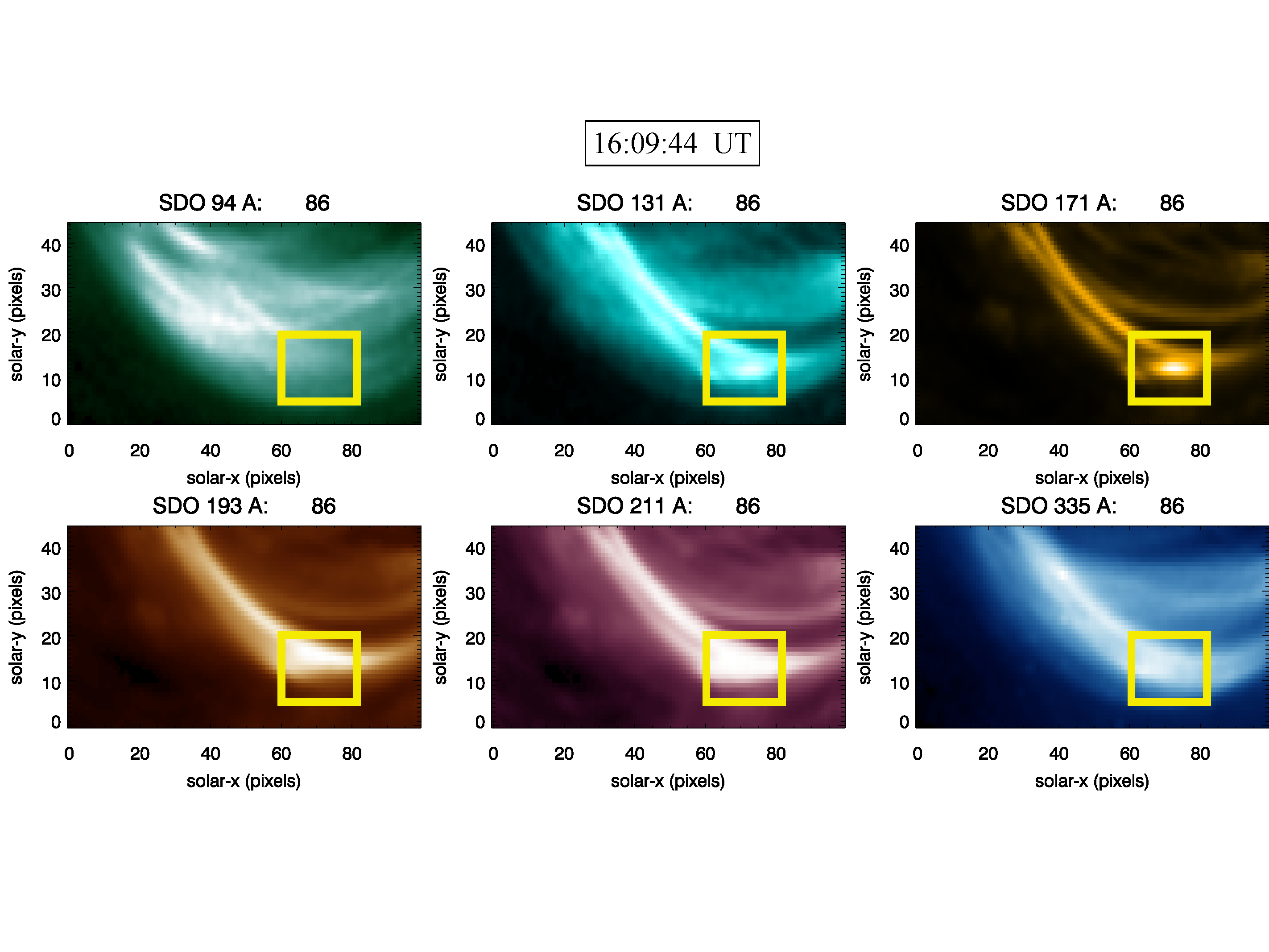}
\caption{All AIA coronal spectral images of 9.4~nm, 19.3~nm and 17.1~nm ({\it top row}) and 19.3~nm, 21.1~nm and 33.5~nm ({\it bottom row}), are presented for time frame no. 86 (16:09:44~UT). This time corresponds to the earliest detection of the coronal rain source within the EUV channels, as determined from the time-distance diagrams. This bright coronal rain source is co-located within the yellow box at the post-flare loop-top, as was previously presented for the bright chromospheric source in Fig.~\ref{fig4} for H$\alpha$ and Fig.~\ref{fig5} for Ca~{\sc ii}, which first appears at 16:10:47~UT.}
\label{fig11}
\end{figure*}


\par In Fig.~\ref{fig10b}, we present the continued evolution of this initial rain source from the loop-top and then along the loop-leg in H$\alpha$ and 30.4~nm, after its appearance as presented in Fig.~\ref{fig10a}. Fig.~\ref{fig10b} reveals the continued spatial correspondence of the flow of the rain for H$\alpha$ and 30.4~nm signatures. As the H$\alpha$ signature of the rain transitions from emission to absorption away from the loop-top, similarly, we detect a reduction of the intensity in the 30.4~nm component within the time range. 

\par At a later time, we detect further evidence for the co-location of the spatially evolving rain strand signatures within different spectral channels, highlighting again, the multi-thermal nature of coronal rain, as shown in Fig.~\ref{fig10c}. Here we present the temporal evolution (top row to bottom row) along the loop-leg, whereby the loop-top is at the bottom-right corner and the foot-point is at the top-left corner of each panel. The trajectory of the loop-leg is very bright and well-defined within the 19.3~nm image at 16:09:05~UT (panel-{\bf A}), where the plasma is expected to be at least 1-2~MK. After 59~s, we have a strong component of the same appear in emission in the 17.1~nm image (panel-{\bf B}). After an additional 55~s, we detect a clear signature of a localised coronal rain source in the 30.4~nm image (panel-{\bf C}) and a subsequent detection within the co-temporal H$\alpha$ red-wing image in the same contoured region from 30.4~nm. In the next time frame, we observe a clear extension of this rain source along the same 19.3~nm loop-leg in both 30.4~nm (faint and elongated) and H$\alpha$ (dark and extended) in the direction of the foot-point from 16:12:23~UT to 16:13:22~UT. We find strong evidence to suggest the presence of a multi-thermal flow in flare-driven coronal rain strands. Given the time cadence between successive rows in Fig.~\ref{fig10c}, we expect that the full half-length of the loop under investigation completely transitions from being hot and dense to having both hot and cold components with local regions of higher density in less than 4~mins. Of course, at the beginning of the formation of the rain strand (in the rain source) this must be taking place at a faster rate. We expect that the catastrophic cooling process occurs during this time window.




\section{REGULARISED SDO/AIA DEM INVERSIONS}

\par In order to understand the evolution of multi-thermal components of coronal rain, we must investigate in more detail the temperature contributions to the AIA passbands. We can then measure the temperature evolution from the X-ray band (as measured with GOES in soft-Xray channels) through to the EUV channels to finally become detectable as coronal rain strands at chromospheric temperatures.  

\par To interpret the temperature contributions within the AIA FOV for each EUV channel in flaring conditions, we adopt a novel approach to calculating the DEM, through applying a generalised regularised inversion procedure, as developed by Hannah \& Kontar (2011, 2012). The DEM ($\xi(T)$) quantifies the amount of plasma emitting within a certain temperature range and relates to the electron density of the plasma ($n_{e}$) as $\xi(T)\,=\,n_{e}^{2}\,dV/dT$, where $V$ is a volume element which will be determined from the pixel area of the EM maps and an atmospheric column depth estimation, $T$ is the plasma temperature. We can investigate the DEM per pixel within a given temperature bin across the FOV of the post-flare arcade in order to: a) investigate the temperature variations during the formation of the coronal rain source and b) extract information about catastrophic cooling leading to rain formation from analysis of the local density variations and subsequent effects on the radiative cooling timescale, addressed further in the discussion section.

\par \citet{2012A&A...539A.146H} constructed a model independent regularization algorithm that makes use of general constraints on the overall form of the DEM vs temperature distribution, in this case in flaring conditions. Assuming optically thin emission in thermal equilibrium (we consider the post-flare phase beyond flare heating leading to non-thermal equilibrium) the DEM is related to the observed dataset $g_{i}$ (observed intensity per AIA spectral channel ($i$) per pixel) as

\begin{equation}
\mathord{g_{i}} = 
\int_{T}\,\mathord{K_{i}}\xi(\mathord{T})dT + \delta\mathord{g_{i}},
\label{DEM}
\end{equation}

\par where $\delta\mathord{g_{i}}$ is the associated error on $g_{i}$ and $K_{i}$ is the temperature response function (for AIA in this case). This ill-posed inversion problem needs to be stabilised. To do so the algorithm introduces extra information by way of a smoothness condition on the source function. By making a prior assumption of the smoothness factor and inverting the data with regularisation \citep{MR0162377}, the algorithm can reliably infer physically meaningful features, which are otherwise unrecoverable from other model dependent approaches such as forward fitting (refer to \citet{2004SoPh..225..293K} for more information). This method has the added advantage of providing errors on the temperature bins through estimation of confidence levels when directly calculating the derivatives and then smoothing the solution to return the DEM. The regularised inversion directly solves the minimisation problem, relating the data set to the expected CHIANTI \citep{1997A&AS..125..149D,1999A&AS..135..339L} DEM model. Here we assume a CHIANTI DEM model for flaring conditions. Using these constraints the inversion problem outlined in Equation.~\ref{DEM} can be restated as

\begin{equation}
\parallel\tilde{\mathbf{K}}\xi(T) - \tilde{\mathbf{g}}\parallel^{2} + \,\lambda\parallel\mathbf{L}(\xi(T) - \xi_{0}(T))\parallel^{2} = \text{min},
\label{reg}
\end{equation}

\par where the tilde represents normalisation by the error, $\lambda$ is the regularisation parameter, $\mathbf{L}$ is the constraint matrix and $\xi_{0}(T)$ is a possible guess solution. We refer to \citet{2012A&A...539A.146H} for a full description of the method which has been applied in many cases involving active region coronal loop systems as observed with SDO \citep{2011ApJ...732...81A,2011ApJ...734...90W,2011ApJ...729L...8F,2011ApJ...727L..52R,2012A&A...539A.146H}. The regularisation method has been very successfully applied to flare studies using RHESSI data for electron flux and density distribution reconstruction as outlined by \citet{2004SoPh..225..293K} \& \citet{2005SoPh..227..299K}.  

\par
\begin{figure}[ht]
\centering
\includegraphics[clip=true,trim=0cm 5cm 0cm 5cm, scale=0.26, angle=0]{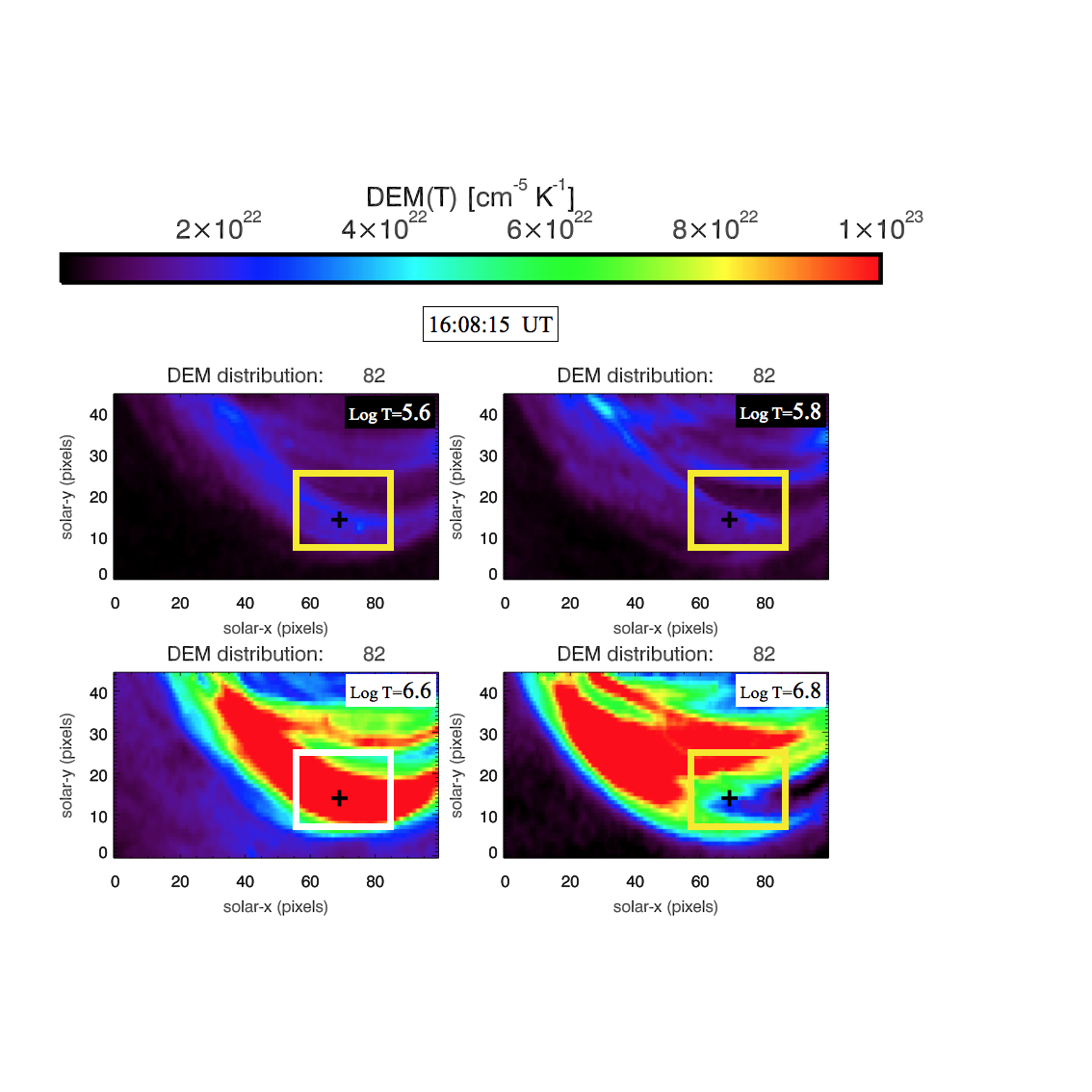}
\caption{The output from the DEM inversion code for calculating the EM in each pixel across the FOV, {\bf before} the formation of the EUV coronal rain loop-top source, i.e. at 16:08:15~UT (frame no. 82). The EM is plotted as a function of increasing temperature in each panel from top left to bottom right. The top 2 panels detail the amount of plasma that is in emission with a temperature {\it below} 1~MK. The bottom 2 panels detail the amount of plasma that is in emission with a temperature {\it above} 1~MK up to Log$_{10}$~T~=~6.8 (6.3~MK). The yellow box is overlaid for context as the location of the chromospheric component of the coronal rain source at the loop-top. The black cross highlights the location of the coronal rain source within the yellow box from H$\alpha$ and Ca~{\sc ii} which will be investigated in detail in Fig.~\ref{fig16}.}
\label{fig13}
\end{figure}

\begin{figure}[h]
\centering
\includegraphics[clip=true,trim=0cm 5cm 0cm 5cm, scale=0.26, angle=0]{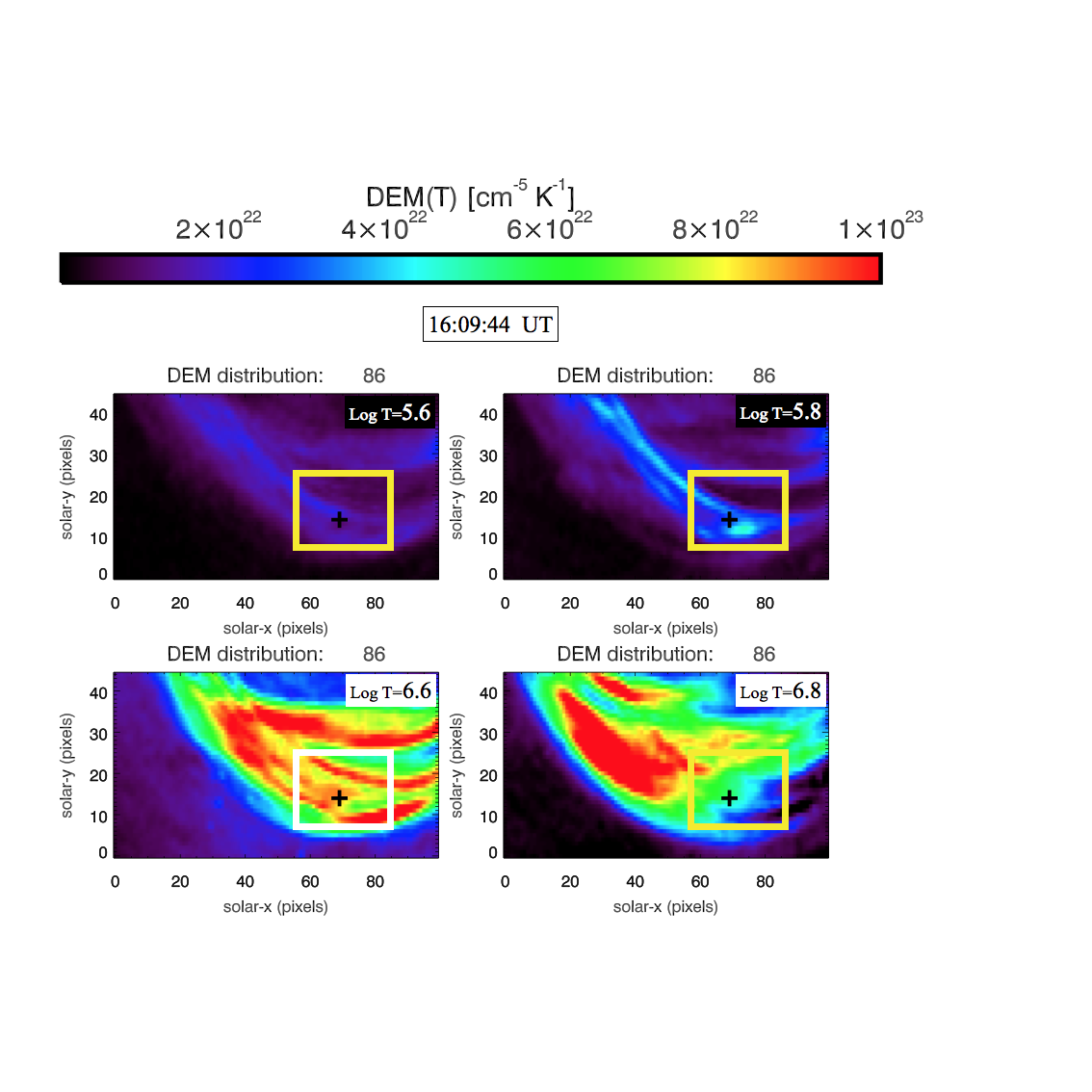}
\caption{The output from the DEM inversion code for calculating the EM in each pixel across the FOV, {\bf at the time of} the formation of the EUV coronal rain loop-top source, i.e. at 16:09:44~UT (frame no. 86). The same format applies as for Fig.~\ref{fig13}.}
\label{fig14}
\end{figure}

\begin{figure}[h]
\centering
\includegraphics[clip=true,trim=0cm 5cm 0cm 5cm, scale=0.26, angle=0]{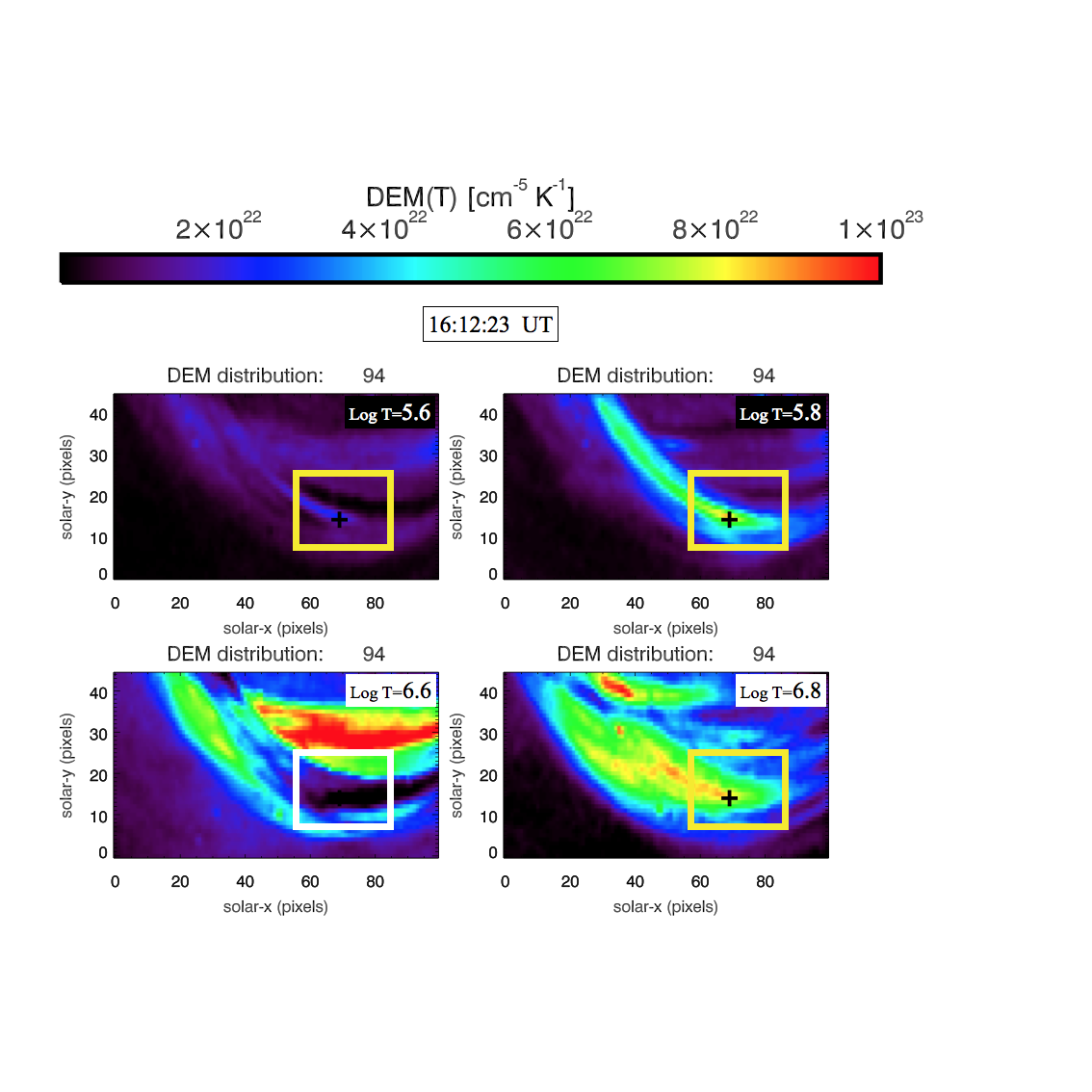}
\caption{The output from the DEM inversion code for calculating the EM in each pixel across the FOV, {\bf after} the formation of the EUV coronal rain loop-top source, i.e. at 16:12:23~UT (frame no. 94). The same format applies as for Fig.~\ref{fig13}.}
\label{fig15}
\end{figure}

 The regularised inversion approach adopted here for calculating DEM's for the observed AIA dataset, will enable an accurate interpretation of the plasma temperature and density variations at the coronal rain source, since information from all spectral channels is incorporated. It is highly likely from our analysis of the EUV data, that the coronal rain source is multi-thermal and so a multi-temperature analysis of the source will help to interpret how the thermal components within the rain source evolve in time and spatially across the FOV. In Fig.~\ref{fig11}, we present the data input from each spectral channel at the time of the rain formation (i.e. frame no.~86). In all of the AIA channels presented, we can see clearly the formation of the post-flare arcade in all spectral channels. Most notably, in the 17.1~nm image we can detect the spatially correlated bright coronal rain source that was revealed within the AIA time-distance diagrams of Fig.~\ref{fig9} (i.e. the yellow boxed region). In Figs.~\ref{fig13}-~\ref{fig15}, the top row panels represent the emitting plasma contribution at below 1~MK and the bottom row represents the plasma DEM for temperature contributions greater than 1~MK, across the FOV.  

\par From Fig.~\ref{fig13}, we sample the EM before the formation of the coronal rain in frame no.~82 (corresponding to 16:08:15~UT). At this time, the post-flare arcade is filled with plasma in emission at greater than or equal to 1$\times$10$^{23}$~cm$^{-5}$~K$^{-1}$ (bright red structures) in the Log~T= 6.6 - 6.8 temperature range. As expected, there is no significant emission within the FOV below 1~MK at this stage in the post-flare decay phase. The EUV post-flare arcade has formed after the decay of the X-ray signal from GOES at this time (see Fig.~\ref{fig2}), whereby the X-ray flux has returned to background levels. 

\par In Fig.~\ref{fig14}, we detect a notable decrease in the DEM in the loop (i.e. within the yellow boxed rain source region) at 16:09:44~UT. The post-flare arcade now contains more patchy red regions and reduced EM per pixel in the Log$_{10}$~T~=~6.6-6.8 temperature range, corresponding to a DEM of 6-7$\times$10$^{22}$~cm$^{-5}$~K$^{-1}$. We expect that catastrophic cooling is established / ongoing in this time prior to its first appearance in the chromospheric channels 35~s later (16:10:19~UT). In Fig.~\ref{fig14} top-right panel, we detect the faint outline in the DEM map of the sub-million Kelvin post-flare loop and associated EUV coronal rain source, which extends towards the foot-point of the flare ribbon at [30,45] in solar-{\it x}/{\it y}. This loop signature in TR plasma follows the trajectory of the H$\alpha$ rain flows as defined by the blue curve cross-cuts in previous figures. 

\par In Fig.~\ref{fig15}, we present the cooling of the loop 159~s later (16:12:23~UT) and the EM has increased in the TR plasma, as the H$\alpha$ rain flow extends from the loop-top to the loop-leg. At this stage, the DEM at the loop-top is now at background levels with respect to plasma at temperatures of Log$_{10}$~T~=~6.6 (4 MK) and at the same time the DEM at Log$_{10}$~T~=~5.8 as increased significantly to 6$\times$10$^{22}$~cm$^{-5}$~K$^{-1}$ and the post-flare arcade is clearly visible at temperatures below 1~MK. Next, we will investigate the DEM vs. temperature profiles associated with the source of the coronal rain formation within the yellow box, as marked by the black cross in Figs.~\ref{fig13},~\ref{fig14},~\&~\ref{fig15} to understand the evolution of the loop cooling process in more detail.


\begin{figure*}[!ht]
\centering
\includegraphics[clip=true,trim=0cm 0cm 0cm 0cm, scale=1.3, angle=0]{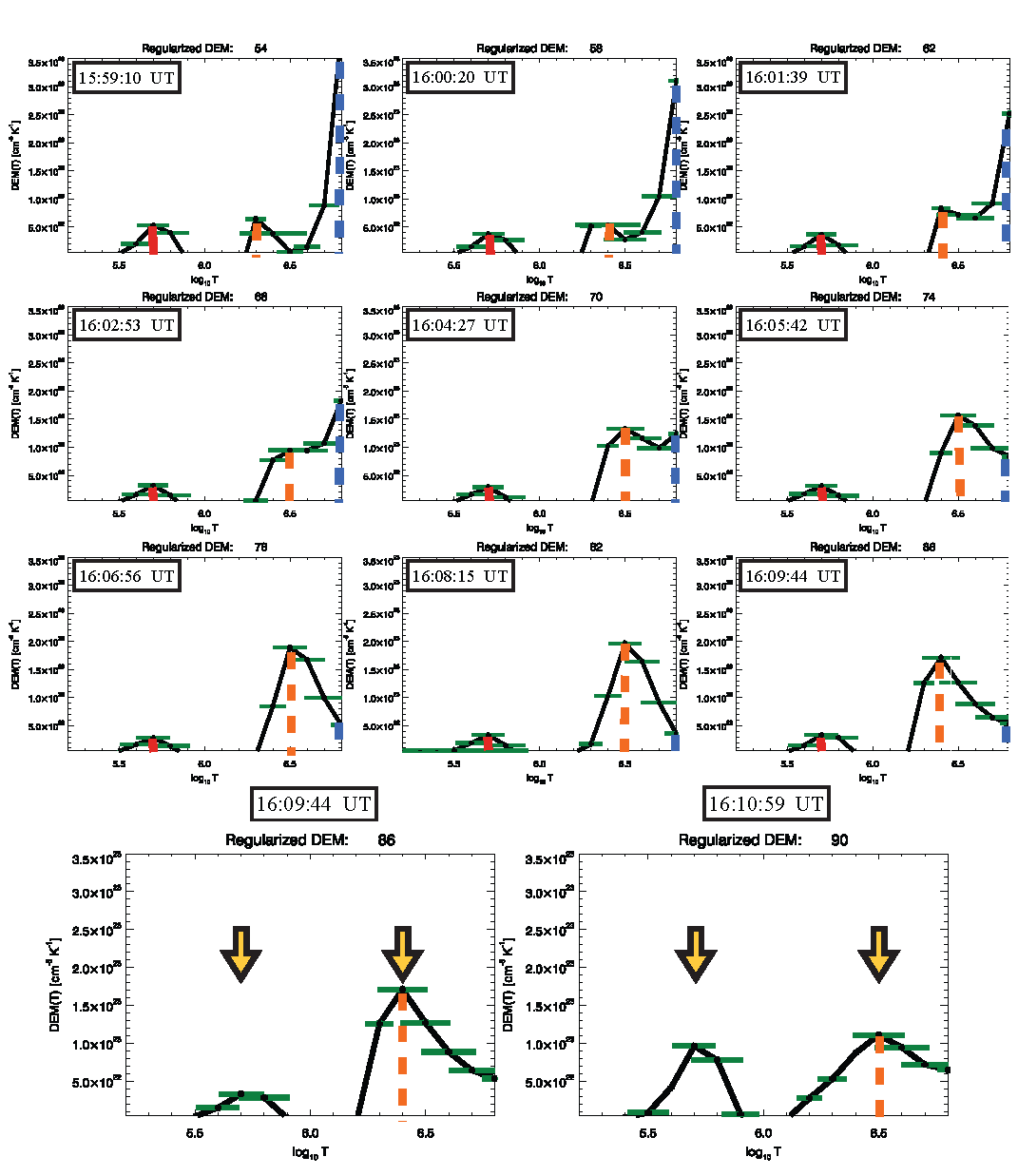}
\caption{A group of 5 pixels from within the black cross of Figs.~\ref{fig13}-\ref{fig15} each contain a DEM vs temperature profile which is averaged and reproduced in each panel here. The evolution of the temperature profile is demonstrated, i.e. from the top-left panel (frame no.~54 - 15:59:10~UT) to the bottom-right panel (frame no.~90 - 16:10:59~UT) and describes the conditions of the post-flare loop-top until the formation of the rain source. The vertical blue dashed line describes the evolution of the hottest detectable plasma emitting at 6.8~MK. The vertical orange dashed line tracks the evolution of the second peak which corresponds to coronal plasma temperatures. The vertical red solid line describes the evolution of the plasma emitting at 10$^{5.7}$~K. Each temperature bin has a measured temperature error and is over-plotted in green. The yellow arrows highlight the relatively large changes in the EM between frame no.~86 and 90 indicating a transition from emission predominantly at coronal temperatures to TR temperatures. The error in DEM measurement at each temperature bin is $\pm$3$\times$10$^{22}$~cm$^{-5}$~K$^{-1}$.}
\label{fig16}
\end{figure*}



\begin{figure*}[!ht]
\centering
\includegraphics[clip=true,trim=0cm 5cm 0cm 7cm, scale=0.49, angle=0]{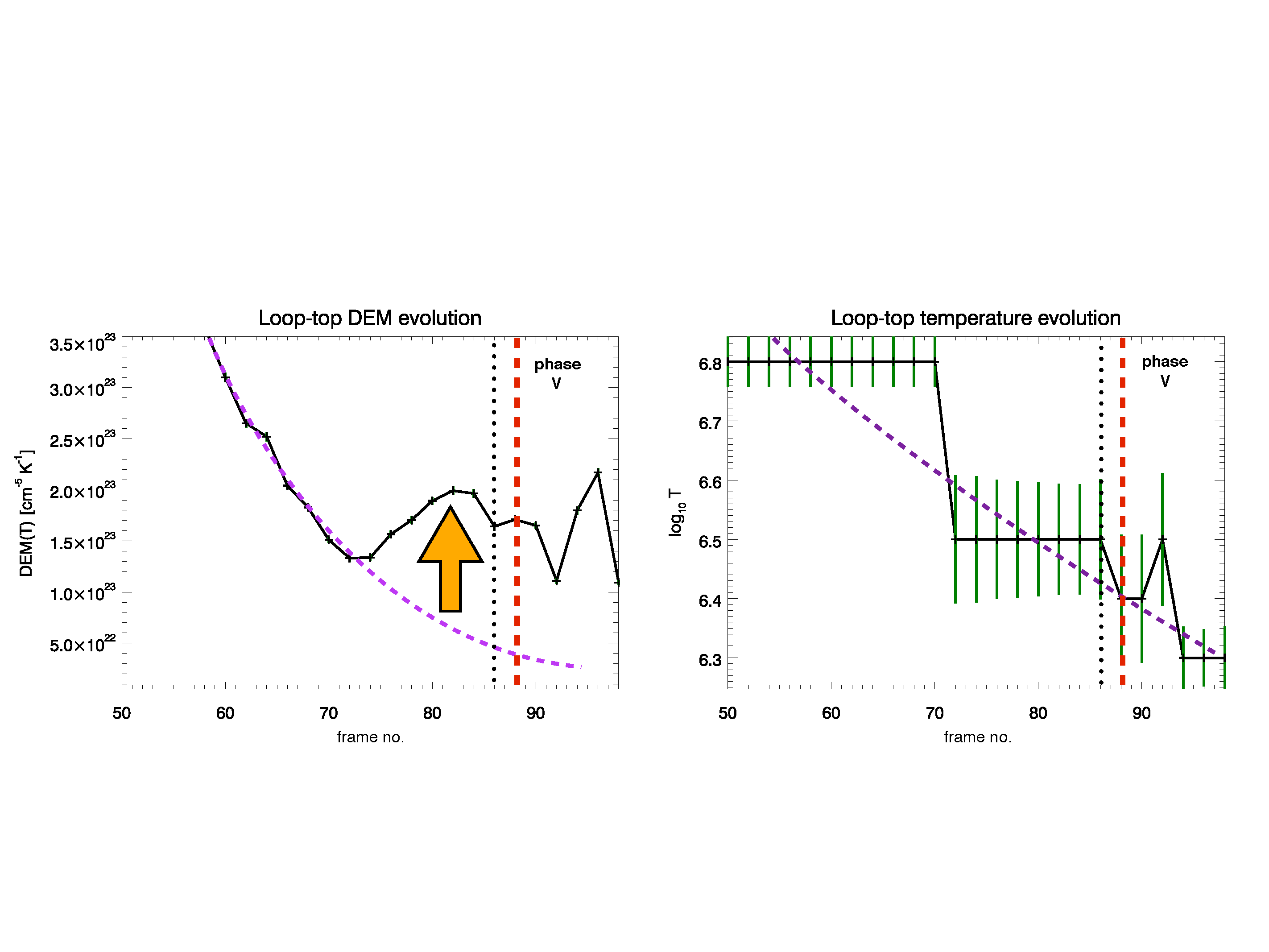}
\caption{Through tracking the peaks in the DEM vs temperature profiles of Fig.~\ref{fig16} we reproduce the evolution of the temperatures ({\it right}) of the rain source in the EUV coronal plasma and corresponding coronal plasma DEM ({\it left}), with respect to frame no. The error in each temperature bin is over-plotted in green. The vertical dotted line (at frame no. 86 at 16:09:44~UT) defines the start of the catastrophic cooling phase (i.e. phase~{\bf V}). The notable EUV EM increase demarcates the onset of cooling through to coronal plasma temperatures (from 7~MK to 1.5~MK), as shown with the upward orange arrow. The red vertical dashed line corresponds to frame no. 88 (i.e. 16:10:19~UT) marking the first appearance / formation of the chromospheric component of loop-top coronal rain. The purple dashed line is a 2$^{nd}$ order polynomial fit to these distributions.}
\label{fig17}
\end{figure*}


\par In Fig.~\ref{fig16}, the DEM vs. temperature profiles within the rain source evolve in time from top-left to bottom-right. In general, we reveal the migration of all DEM peaks from right to left, i.e. leading to greater EM for progressively lower temperature plasma in time. For instance, between 15:59:10~UT and 16:06:56~UT the EM in the highest temperature bin (i.e. having a peak at Log$_{10}$~T~=~6.8 and marked with the vertical blue dashed line) reduces extensively in DEM. Simultaneously, we detect a progressive increase in the EM in the temperature range of Log$_{10}$~T~=~6.5 (as marked with the vertical orange dashed line). Hence, within 466~s a large component of the plasma temperature at the source has dropped by $\sim$3.1~MK, which corresponds to at least 6500~K~s$^{-1}$. Later, between 16:08:15~UT and 16:09:44~UT this plasma temperature peak, as marked with the orange dashed line, appears to migrate from Log$_{10}$~T~=~6.5 to Log$_{10}$~T~=~= 6.4, which corresponds to an even faster rate of cooling of 7300~K~s$^{-1}$. The yellow arrows ({\it bottom row} of Fig.~\ref{fig16}) highlight the relatively large changes in the DEM at the coronal rain source when plasma emitting at greater than 1.5~MK (Log~T = 6.2) at frame no.~86 reduces in EM while plasma at $\sim$0.5~MK substantially increases by frame no.~90 (16:10:59~UT). Notice the peak at Log$_{10}$~T~=~5.7 (corresponding to TR plasma emission) has approximately the same emitting contribution as the peak at Log$_{10}$~T~=~6.2, indicating that most of the plasma is cooling to sub-million Kelvin temperatures. This process occurs during the first appearance of chromospheric component of the rain in its formation, supporting the argument that the source is indeed multi-thermal, as well as, being very thermally active. By tracking the migration of the DEM peaks in the DEM vs. temperature profiles at the rain formation site, we can interpret the temporal evolution of temperature (see Fig.~\ref{fig17} {\it right}) and EM (see Fig.~\ref{fig17} {\it left}) at the source. 

\par In Fig.~\ref{fig17} ({\it left}), we discover the sudden increase in EM from frame no.73 (i.e. 16:05:26~UT) until frame no.~82 (i.e. 16:08:15~UT). This increase corresponds to the response of the coronal plasma increasing in emissivity in the temperature range of 1.5-7~MK. In Fig.~\ref{fig17} ({\it right}) confirms that during this time interval we detect a continuous cooling of the coronal plasma. The large orange arrow marks the time of the peak in this period of increased EM of coronal plasma which occurs at frame no.~82 (16:08:15~UT). Next, within 4 frames (89~s) until 16:09:44~UT, the cooling into the 17.1~nm passband becomes detectable at the loop-top source from imaging, i.e. the first signature of the local rain formation present in Figs.~\ref{fig10a},~\ref{fig10c}-(panel~B),~\ref{fig11}~\&~\ref{fig15}. In Fig.~\ref{fig17}, the commencement of the TR plasma emission, at below 1~MK, is marked with a black dotted line which we define as the start of catastrophic cooling, referred to herein as phase~{\bf V}. Between frames 82 and 86 the EM has started to decrease in the coronal loop-top plasma and cooling starts to become dominant in TR plasma. Until 16:10:19~UT (i.e. an additional 35~s from frame no.~86), we have a short period of accelerated cooling to chromospheric temperatures at this source, leading to the appearance of the source in H$\alpha$ in emission followed by absorption and this interval coincides with a further increased contribution from plasma in emission at $\sim$0.5-1~MK in the loop-leg. In summary, the time interval through which catastrophic cooling occurs, when the temperature at the loop-top source drops by $\sim$1.5~MK, is greater than 35~s and less than 124~s, at the start of the decline the corona plasma EM peak (see orange arrow Fig.~\ref{fig17} {\it left}) and first appearance of the chromospheric component of the rain (see red dashed line Fig.~\ref{fig17} {\it right}). 

  


\par In order to place the catastrophic cooling process leading to coronal rain formation in the context of post-flare loop cooling, we have appended the GOES, AIA and CRISP temperature profiles into a summarising cooling curve in Fig.~\ref{fig18}. This cooling curve uniquely connects the hottest components at the flare temperature peak with the formation of the chromospheric component of the coronal rain strands at the loop-top and shortly after in the loop-legs. From Fig.~\ref{fig18}, we reveal 5 phases in the cooling process, as bounded by the vertical dashed lines and these regions are banded in different colours. The physical interpretation of this summarising cooling curve will form the basis of the discussion section, as well as, providing new insights into the earliest onset of catastrophic cooling in the EUV prior to its appearance at chromospheric temperatures.

\section{DISCUSSION}

\par We reveal in detail a clear association between flare sources of chromospheric evaporation and the subsequent cooling of the heated plasma, which falls back to the same source as chromospheric coronal rain (see Fig.~\ref{fig3}). Dense clumps / strands of partially ionised chromospheric coronal rain, flow in a multi-thermal stream along a trajectory prescribed by the post-flare magnetic arcade. Subsequently, the chromosphere-corona mass cycle is investigated at the highest achievable resolution using GOES, SDO / AIA and SST / CRISP instruments covering a broad spectral range spanning soft X-rays to the (E)UV, near-IR and visible wavelengths. 


\begin{figure*}[!ht]
\centering
\includegraphics[clip=true,trim=0cm 0cm 0cm 0cm, scale=1.1, angle=0]{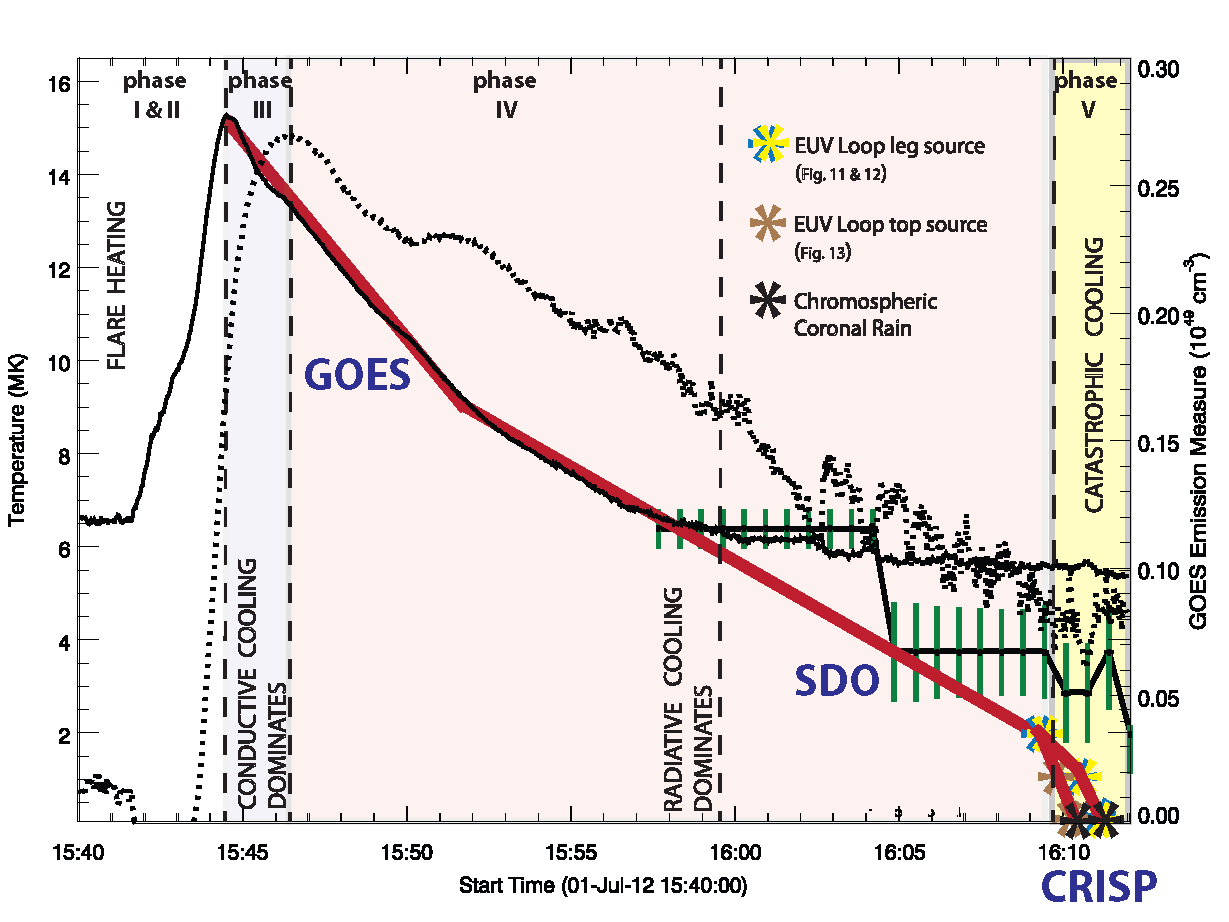}
\caption{The cooling curve (red solid curve) extends from the temperature peak of the flare impulsive phase observed in GOES (solid black curve), through the EUV as observed with AIA (solid black lines with green error bars) until the start of the formation of coronal rain strands as observed with CRISP (star symbols). The temperature profile has its axis on the LHS. During the SDO / EUV stage the cooling curve connects the start of the most prominent changes in the temperature evolution of the DEM vs temperature profiles (see Fig.~\ref{fig16}). Five phases ({\bf I-V}) describe the evolution of the dominant cooling processes, i.e. in accordance with \citet{2007A&A...471..271R} and these intervals are labelled and coloured. The transition times for each respective cooling mechanism corresponding these phases are calculated using the Cargill Model and presented with vertical dashed black lines. Phase~{\bf I-II} represents the heating / onset of the flare, phase~{\bf III} represents the dominant conductive cooling phase, phase~{\bf IV} represents the transition to radiative cooling including the onset of the dominant radiative cooling phase (marked with vertical dashed lines) and finally phase~{\bf V} represents the start of catastrophic cooling. During catastrophic cooling the red cooling curve splits into two branches, corresponding to loop-top coronal rain temperature evolution (brown symbols) and loop-leg coronal rain temperature evolution (yellow symbols), respectively. The black dashed curve is the GOES EM with axis on the RHS.}
\label{fig18}
\end{figure*}


\par During the flare impulsive phase, the plasma temperature peaked at $\sim$15.4~MK, via direct heating of the chromosphere resulting in chromospheric evaporation \citep{1976SoPh...49..359A,1976SoPh...50...85K}. Chromospheric evaporation can be driven by a number of  heating mechanisms, aside from non-thermal particle beams \citep{2015ApJ...813..113B} and it can be classified as either explosive or gentle \citep{1985ApJ...289..425F,2006ApJ...638L.117M,2006ApJ...642L.169M}. Thermal conduction from the corona can drive the expansion of hot, dense chromospheric material into post-flare arcades, however, eventually it is expected that the conductive heat flux will no longer compensate for the radiative losses in the corona and the loops will rapidly begin to cool \citep{1999ApJ...512..985A,2001ApJ...553L..85K}. Radiative losses increase and dominant over conductive losses and  at a later stage a loop-top thermal instability leads to catastrophic cooling (i.e. accelerated cooling) to chromospheric temperatures and we observe a substantial loop drainage / depletion, as observed here in H$\alpha$. During the thermal instability, the decrease in temperature in the loop is accompanied by a decrease in pressure, which then accretes plasma from the surrounding atmosphere leading to the localised formation of dense rain condensations \citep{1971SoPh...19...86G,1974SoPh...35..123H,1991ApJ...378..372A,2004A&A...424..289M,2013ApJ...771L..29F}. Clumpy condensations eventually become dense enough to fall under gravity back to the surface (overwhelming the opposing magnetic pressure force of the loop arcade) resulting in a catastrophic depletion of plasma in the loop. The post-flare loop cooling through the soft X-ray to EUV channels leading to the formation of localised dense clumps of multi-thermal coronal rain in H$\alpha$, first in emission then in absorption, has not been observed until now.

\par  To do so, we employ the hydrodynamic 0D model by \citet{1995ApJ...439.1034C}, which describes the cooling of post-flare loop plasma during the flare decay phase and can be used to interpret the timescales associated with the cooling curve of Fig.~\ref{fig18}. After a detailed statistical analysis, it was found that the Cargill model provides a very well defined lower limit on flare cooling times and that radiation is the dominant loss mechanism throughout the cooling for 80\% of flares \citep{2013ApJ...778...68R}. For the remaining 20\%, conduction dominates initially, before cooling becomes dominated by radiation. For simplification, we refer to 0D models such as the Cargill model, that assign field-aligned averages of the hydrodynamic properties of cooling loops, within the limits of optically thin plasma at temperatures $>$1-2~MK. Field-aligned averages are justified by the fact that hydrodynamic properties (such as temperature, pressure and density) do not vary much along the length of the coronal loop within the corona, except near the interface of the corona and TR, which is characterised by steep gradients. The 0D assumptions are considered to be acceptable in comparison with 1D models \citep{2008ApJ...682.1351K}. If the cooling curve from Fig.~\ref{fig18} can be explained by known energy loss mechanisms, we should expect that the onset of catastrophic cooling should occur within the expected timescale for radiative cooling. Next, we will calculate the energy loss rates due to conduction, radiation and enthalpy-based radiative cooling, through considering the energy transport equation in cooling flare loops. 

\par In order to simplify the problem of calculating the cooling rates in post-flare loop arcades, the Cargill model assumes that the plasma is confined to the axis of the magnetic field ($s$), so there is unsubstantial cross-field diffusion of energy (i.e. strands of plasma are thermally isolated), which is an acceptable assumption under solar conditions, given the relatively short Debye length scales compared with the Larmour radius for particle collisions (NB fast thermal modes produced in numerical models of coronal rain can leak energy across fields). In this study we assume that field aligned thermal conduction is dominant since that helps to explain why we observe very clearly defined plasma strands in loop arcades. Furthermore, we assume a single fluid approximation and do not consider the collisional energy loss rate between different particle species. The energy transport equation can then be written of the form as described in \citet{2013ApJ...778...68R} as

\begin{equation}
\frac{1}{\gamma - 1}\frac{\partial\,p}{\partial\,t}  = -\,\frac{1}{\gamma - 1}\frac{\partial}{\partial\,s}(p\,u_{s})\,-\,p\,\frac{\partial\,u_{s}}{\partial\,s}\,+\,\frac{\partial}{\partial\,s}\,F_{c}\,-\,n_{e}^{2}\,\Lambda\,(T)\,+\,h,
\label{energyequation}
\end{equation}

\par where $\gamma$ is the adiabatic constant, $p$ is pressure, $u_{s}$ is the plasma flow velocity along the axis of the magnetic field ($s$), $F_{c}$ is the conductive heat flux, $n_{e}^{2}\,\Lambda$($T$) is the radiative loss rate (assuming optically thin conditions) and $h$ is the heating rate per unit volume. The 1$^{st}$ and 2$^{nd}$ terms on the RHS of Equation~\ref{energyequation} represents energy losses due to flows within the cooling loop, which is associated with enthalpy-based radiative cooling, as well as, expansion and contraction of the loop. Here we assume that the loop does not appreciably expand or contract during the decay phase therefore assumptions about the loop volume and length at the start of cooling are the same as those applied at the latest stages of cooling. It should be noted that the radiatively dominant phase is a result of energy loss not just by radiation but also by an enthalpy flux of mass flows falling down the loop-legs \citep{2008A&A...486L...5B}. Enthalpy-based cooling becomes significant at the latest stages of flare loop cooling and the enthalpy flux is considered to be important in balancing the TR emission / radiative losses \citep{2010ApJ...717..163B,2013ApJ...772...40C}. The third term on the RHS, defines the conductive losses from the divergence of the heat flux, $F_{c}$, which is defined as $F_{c}\,=\,-\,\kappa\frac{\partial\,T}{\partial\,s}$ and $\frac{\partial\,T}{\partial\,s}$ is the temperature gradient. Here $\kappa$ is the Spitzer thermal conductivity (i.e. $\kappa\,=\,\kappa_{0}\,T^{\frac{5}{2}}$), such that $\kappa_{0}\,\approx\,10^{-6}$. The fourth term on the RHS represents the radiative loss rate, where $\Lambda\,(T)$ is the optically thin radiative loss function. A scaling law for $\Lambda(T)$ can be assumed (see review by  Reale \& Landi, 2012), using the parameterisation of \citet{1978ApJ...220..643R}, resulting in $n_{e}^{2}\,\Lambda\,(T)\,=\,n_{e}^{\zeta}\,\chi\,T^{\alpha}$. 

\par The evolution and energy transport in the post-flare arcade plasma is controlled by the balance between conductive and radiative losses, together with flows and decay phase heating processes ($h$), such as additional magnetic reconnection. Next, we will consider the cooling timescales due to each of these processes, independently, using the derived properties from the observations.

\subsection{Conductive cooling}

\par \citet{2007A&A...471..271R} defined 4 phases describing the evolution of confined plasma in flare loops. Here we characterise this flare along those guidelines and introduce a phase {\bf V} for catastrophic cooling. Phases {\bf I} \& {\bf II} describe the {\it heating} and {\it evaporation} of the plasma from the start of the flare heat pulse in the corona to the temperature peak $T$. The heat pulse is efficiently conducted to the cool chromospheric plasma which is strongly heated and expands filling the loop with hot dense plasma. This results in a rapidly increasing emission measure offset from the temperature peak. Phase {\bf III} corresponds to the {\it conductive cooling phase} and occurs between the end of heating and the EM peak by efficient thermal conduction \citep{2004ApJ...605..911C}. Conductive cooling may account for faster cooling timescales in flaring conditions \citep{1982ApJ...258..373D} when the temperature gradients are largest. The conductive cooling phase starts at the temperature peak and continues to the peak of the EM \citep{1994ApJ...422..381C} which is common in flare observations \citep{1993A&A...267..586S}. In this flare, at the start of the cooling phase {\bf III}, the temperature peaks measurably before the EM (see Fig.~\ref{fig18}). The conductive cooling timescale ($\tau_{c}$) can be calculated by neglecting heating, radiative losses and energy transport due to flows terms. The model can be further simplified through assuming that the plasma is isothermal and obeys the ideal gas law, whereby  $p$=$\frac{3}{2}n_{e}\,k_{B}\,T$, where $k_{B}$ is the Boltzmann constant. Therefore, Equation~\ref{energyequation} becomes

 \begin{equation}
\frac{\partial}{\partial\,s}\,[\kappa_{0}\,T^{5/2}\frac{\partial\,T}{\partial\,s}]  = \frac{k_{B}}{\gamma - 1}\,n_{e}\,\frac{\partial\,T}{\partial\,t}
\label{energyequation2}
\end{equation}

\par After integration of Equation~\ref{energyequation2} one can derive an approximated relationship for the conductive cooling timescale \citep{1995ApJ...439.1034C} as

\begin{equation}
\tau_{c} = 4\,\times\,10^{-10}\frac{n_{e}L^{2}}{T^{5/2}} 
\label{conductive_timescales}
\end{equation}


\par During phase~{\bf III}, starting at 15:44:30~UT when the temperature peaks at 15.4~MK (according to GOES lightcurve), conductive losses dominate and we can calculate the conductive loss timescale (using Equation~\ref{conductive_timescales}). From Fig.~\ref{fig18} at the temperature peak, we measure the EM of 0.165$\,\times\,10^{49}$~cm$^{-3}$. In order to calculate the electron density from the GOES EM we need to assume a volume of emitting plasma. Here we assume a volume of a cylinder describing the soft X-ray post-flare loop with cylinder length equal to 2$L$ and diameter of 10~arcsec, which is determined from the observations of the loop cross-sections in AIA at a later stage in the cooling process (i.e. see inset panel of Fig.~\ref{fig2}). Taking the observed loop half length ($L$) from the observations as $L$=$\,3.2\,\times\,10^{9}$~cm, we hereby assume an emitting volume of 2.14$\times$10$^{27}$~cm$^{3}$ and with a filling factor of 1 we estimate an electron density at the start of phase~{\bf III} to be 2.77$\,\times\,10^{10}$~cm$^{-3}$, which is reasonable under flaring conditions. With these estimates we calculate the conductive cooling timescale of $\tau_{c}$ = 122~s, hence, a conductive cooling end time at $\sim$15:46:32~UT which indeed corresponds with the peak in the observed GOES EM. The assumption that the loop volume does not change appreciably between the start of the decay phase and the end is warranted, considering this clear correspondence between the model conductive cooling time and the observed peak in the EM from GOES. At the end of the dominant conductive cooling phase~{\bf III}, the plasma temperature has dropped to 13.4~MK. This corresponds to a cooling rate of $\sim$14300~K~s$^{-1}$. Conductive cooling does not dominate the evolution of the cooling loop at all temperatures. At the end of phase~{\bf III}, the observed EM reaches a peak (as expected) and assuming the emitting volume has not changed significantly within 2~mins, then the electron density will have increased to 3.5$\,\times\,10^{10}$~cm$^{-3}$. Consequently, there will be an increase in radiative losses and eventually a transition to predominantly radiative cooling, as the loop temperature decreases and the loop density increases \citep{1976SoPh...49..359A,1994ApJ...422..381C}. 

\subsection{Dominance of radiative cooling}

\par The efficiency of conductive cooling decreases with the temperature drop off while the efficiency of radiative cooling increases and we enter phase~{\bf IV}, following \citep{2007A&A...471..271R}. The EM can be approximated as a power-law of the form EM$\sim$T$^{b}$ \citep{1993SoPh..144..217D,1996ApJS..106..143B,2011ApJ...740....2W}. We fit a power-law with an index $b$ of 3.6 to this curve which is indeed characteristic of strong heating events such as flares. Next, we consider radiative cooling timescales ($\tau_{r}$) from the Cargill model and compare with the observations. The radiative cooling timescale can be calculated from Equation~\ref{energyequation} through neglecting heating, conductive losses and energy transport due to flow terms resulting \citep{1995ApJ...439.1034C} in

\begin{equation}
n_{e}^{\zeta}\,\chi\,T^{\alpha}  = \frac{k_{B}}{\gamma - 1}\,n_{e}\,\frac{\partial\,T}{\partial\,t}.
\label{energyequation3}
\end{equation}

\par After rearranging the terms and integrating we can express the radiative cooling timescale as

\begin{equation}
\tau_{r}\,=\,\frac{k_{B}}{(\gamma\,-\,1)(1\,-\,\alpha)\chi}\,\frac{T^{1-\alpha}}{n_{e}^{\zeta\,-\,1}}. %
\label{radiative_timescales}
\end{equation}

\par In the limit 10$^{6}$~-~10$^{7}$~K, we expect that $\zeta\,=\,2$, $\chi\,=\,1.2\,\times\,10^{-19}$, $\alpha\,=\,-1/2$ and $\gamma\,=\,5/3$ according to \citet{1978ApJ...220..643R} and after inputting the same plasma temperature and density properties at 15:46:32~UT (i.e. the start of the dominant radiative cooling phase~{\bf IV}), we calculate a radiative cooling timescale of $\sim$4764~s (i.e. $\sim$1~hr 20~mins) which is much longer than the expected radiative cooling timescale from observation (i.e. 26 mins after the temperature peak we have the first appearance of the chromospheric component of coronal rain). We do not assume that there is a discrete start and end time with respect to the dominance of radiative cooling versus conductive cooling but these time estimates give a good indication of approximately when we can expect to detect such a transition. Similarly, \citet{2009A&A...494.1127R} investigated a C-class flare and also found values for $\tau_{c}$ (300~s) to be much less than $\tau_{r}$ ($\sim$4000~s) which may indicate that for relatively weak C-class flares we may not expect a large conductive cooling timescale given that the peak temperatures will be relatively low. A comparison was made between the Cargill approach and the Enthalpy based thermal evolution of loops (EBTEL) model \citep{2008ApJ...682.1351K} during the flare cooling phase. EBTEL simultaneously calculates the conductive and radiative losses throughout the flare and estimates the onset time at which one energy transfer mechanism dominates over the other. In a comparison between the Cargill model and EBTEL it was shown by \citet{2009A&A...494.1127R} that both models where in agreement with respect to the predicted time ($\tau_{*}$) at which radiative losses dominate over conductive losses. According to the Cargill model, the time at which the dominant cooling mechanism changes from conductive to radiative cooling ($\tau_{*}$) can be defined as the ratio of the respective timescales \citep{1995ApJ...439.1034C} as

\begin{equation}
\tau_{*} = \tau_{c_{0}}\left[\left(\frac{\tau_{r_{0}}}{\tau_{c_{0}}}\right)^{7/12} -1\right],
\label{dominancetime}
\end{equation}

\par where subscript "0" denotes the initial values of the conductive cooling timescale at the start of the cooling phase ($\tau_{c_{0}}$) and the radiative cooling timescale at the start of the radiative phase ($\tau_{r_{0}}$). The time $\tau_{*}$ at which $\tau_{c_{0}}\approx\tau_{r_{0}}$, i.e. when the dominant loss mechanism switches from conduction to radiation \citep{2004ApJ...605..911C}, is 913~s which corresponds to $\sim$15:59:42~UT, as marked  with a vertical dashed line in Fig.~\ref{fig18}. The predicted temperature from the start of the cooling phase which is expected at $\tau_{*}$ is given \citep{1995ApJ...439.1034C} as

\begin{equation}
T_{*} = T_{0}\left(\frac{\tau_{r_{0}}}{\tau_{c_{0}}}\right)^{-1/6}.
\label{dominancetemp}
\end{equation}

\par Here we expect the temperature of the plasma in the flare loop to have cooled to 8.4~MK. The observed temperature at $\tau_{*}$ from Fig.~\ref{fig18} is in the range of 6-7~MK. Indeed the Cargill model make an accurate approximation in this regard. According to \citet{1995ApJ...439.1034C} if $\tau_{c_{0}}\ll\tau_{r_{0}}$ then the total loop cooling time ($\tau_{cool}$), assuming conductive cooling is evaporative rather than static, can be approximated as

\begin{equation}
\tau_{cool} \approx \frac{5}{3}\left(\tau_{r_{0}}^{7/12}\tau_{c_{0}}^{5/12}\right).
\label{tcool}
\end{equation}

\par Here we calculate the total loop cooling time as 1848~s which corresponds to an end time of $\sim$16:15~UT. We detect cool, dense chromospheric components in the coronal rain formation, initially at the loop-top, at 16:10:19~UT, which is clearly earlier than the predicted time. The measured densities may be larger than what is reported here, as a result of overestimating the volume of the emitting plasma, which would lead to reduced radiative cooling timescales in the later phases of cooling \citep{2012A&A...543A..90R}. From the averaged red line in Fig.~\ref{fig18} (which extends from the temperature peak to the formation of cool, dense chromospheric rain clumps within phase-{\bf V}) the temperature drop throughout phase~{\bf IV} corresponds to a cooling rate of 7300~K~s$^{-1}$. Next, we consider the properties of the loop cooling into the EUV coronal and TR plasma, using the AIA DEM calculations. From Fig.~\ref{fig18}, we detect a significant steepening of the temperature decrease from phase~{\bf IV} into phase-{\bf V}, which we class as catastrophic cooling which commences prior to the first appearance of the chromospheric rain component. With the red curve we estimate an acceleration in the rate of cooling of the plasma reaching a maximum at 22700~K~s$^{-1}$ between frame no.~86 and~88. This apparent acceleration in the cooling is marked as phase~{\bf V} in Fig.~\ref{fig18} and changes in the properties of the cooling plasma during this phase will give valuable insight into the nature of {\it catastrophic cooling}. 


\subsection{Transition from radiative cooling to catastrophic cooling and rain flows}

\par The sequential appearance of the formation of coronal rain from the EUV, visible and near-IR passbands is imaged and indicates a rapid progression of the plasma temperature through these passbands, as presented in Figs.~\ref{fig10a},~\ref{fig10b}~\&~\ref{fig10c}. In Fig.~\ref{fig16}, we presented the DEM vs temperature profiles in order to understand the nature of the plasma temperature and density properties immediately prior to the onset of catastrophic cooling and at the location of the rain formation. So far we have considered the role of conductive and radiative cooling in the early phases of the decay of the flare. At later stages energy transport in mass flows could come from enthalpy-based cooling (i.e. enthalpy flux). Even though enthalpy flux removes energy from the corona in mass flows it is not an energy loss mechanism like radiation, but rather, it redistributes energy from the corona to the TR \citep{2010ApJ...717..163B}.

\par Whenever the radiative cooling mechanism becomes dominant  loop depletion starts very slowly at first and then progressively faster since the pressure decrease can no longer support the condensing plasma \citep{2007A&A...471..271R}. In this scenario, entalpy-based radiative cooling may play an important role in sustaining the TR radiative losses through a redistribution of the energy driven by downflows. \citet{2008A&A...486L...5B} demonstrated that for certain flow velocities the enthalpy flux (mechanical transport of energy) could balance the radiative energy loss in cooling active region loops in order to avoid catastrophic cooling and, in turn, power the TR radiation. According to \citet{2008A&A...486L...5B}, the critical velocity that the downflow must reach in order to drive an enthalpy flux sufficient to sustain the TR radiative emission and avoid catastrophic loop drainage is in the range of 15.3-76~\kms\ along the loop-leg (close in height to the TR) for loop densities in the range of 1-5$\times$10$^{10}$~cm$^{-3}$ when the loop apex is 1~MK. Mass transport associated with the cooling loop in this study is clearly present in the EUV for coronal plasma, as well as, TR plasma at the end of phase~{\bf IV} and start of phase~{\bf V} when the loop temperature is expected to be between 10$^{5.9}$~K and 10$^{6.2}$~MK (see red curve at start of phase~{\bf V} in Fig.~\ref{fig18} and Fig.~\ref{fig17} {\it bottom row}). Between frame no. 86 and 88, corresponding to 35~s into phase~{\bf V}, we detect the evolution from coronal plasma temperatures ($>$1.5~MK) to radiative losses in the TR plasma ($\sim$0.5~MK) in the DEM profiles localised to the loop-top sources of coronal rain. After this short interval, we detect catastrophic loop drainage originating at the loop-top in multi-thermal mass flows, before appearing in the loop-legs, where we detect apparent velocities of $\sim$54.5~\kms\ (see Fig.~\ref{fig8}). Furthermore, we have calculated the densities in the flare loop of $\sim$3.5$\times$10$^{10}$~cm$^{-3}$, therefore, according to \citet{2008A&A...486L...5B} we should expect that the enthalpy flux will balance the TR radiative losses in the loop-leg. Indeed, during this period of enthalpy flux in the loop-leg, we detect the increased radiative emission signature of a loop in the TR plasma in the DEM maps at 16:12:23~UT with temperatures of 10$^{5.8}$~K (see Fig.~\ref{fig15} {\it top right panel}) where we detect the H$\alpha$ flows. This co-spatial, multi-thermal flow field development leading to bright TR emission signatures along the loop leg between 16:10:59~UT and 16:13:22~UT. is presented in detail in Fig.~\ref{fig10c}. This loop exists at TR plasma temperatures for as long as the high speed mass flow are present, as the enthalpy-based cooling model predicts. 

\par Despite this, we find that catastrophic cooling to chromospheric temperatures at the loop-top source specifically, has rapidly surpassed the mechanism of enthalpy flux in powering the TR losses, originally proposed to prevent collapse and extensive loop depletion. In other words, we have rapid loop-top catastrophic cooling, followed by multi-thermal mass flows, leading to loop-leg enthalpy flux to power the TR radiative losses. This outcome is somewhat supported by \citet{2013ApJ...772...40C} who compared analytical models with numerical results to show that catastrophic cooling is due to the inability of a loop to sustain radiative / enthalpy cooling below a critical temperature, which can be $>$1~MK in flares. It may be interpreted that the enthalpy flux in mass flows at the loop apex, cannot be expected to be as sufficient as along the loop-leg, in sustaining the enthalpy-based radiative cooling process, thereby enabling accelerated cooling at the apex.

\par Next, we investigate the observed catastrophic cooling phase in more detail, using the temperature and DEM changes within Fig.~\ref{fig17} to estimate local changes in the plasma density of the chromospheric component of the rain source, in order to better characterise the onset of catastrophic cooling observed here with respect to the analytical / numerical models of \citet{2012A&A...543A..90R},~\citet{2010ApJ...717..163B}~\&~\citet{2013ApJ...772...40C}.

\subsection{Plasma properties during catastrophic cooling}

\par The DEM profiles presented in this analysis (see Fig.~\ref{fig16} {\it bottom row}), between 16:09:44~UT and 16:10:59~UT, reveal important information regarding the TR plasma properties during catastrophic cooling. Assuming that we have a multi-thermal rain structure forming co-spatially at the loop-top source and assuming the volume of the emitting region as a sphere with the diameter of the contoured region centred (for the longest diagonal) on the rain formation region (see 30.4~nm black contour at 16:10:04~UT from Fig.~\ref{fig10a}), then we can calculate the density of the TR plasma component from the DEM locally at the source of the rain. From this estimate, we can momentarily infer the density of the H$\alpha$ chromospheric component by deducing an ideal gas pressure balance across the structure whereby the rain source has a hotter outer sheath of TR plasma coating a cooler more dense core in emission at 16:10:19~UT in H$\alpha$ (see Fig.~\ref{fig10a}). It is important to note that this determination of plasma properties from the DEM will represent a lower limit to the density of the rain strands before loop depletion, given that condensation will continue to increase the plasma density locally, until it is observed in absorption in H$\alpha$ and proceeds to flow along the loop-legs. Through considering the DEM in the range of 5.5$<$Log$_{10}$~T$<$5.7 between 16:09:44~UT and 16:10:59~UT (see Fig.~\ref{fig16} {\it bottom row}), we notice a significant increase due to the appearance of the rain source and we calculate the EM by integrating the DEM curve across this temperature range as follows

\begin{equation}
\int^{5.7}_{5.5}\,\xi\,(T)\,dT\,=\,EM.
\label{DEMtoEM}
\end{equation}

\par Therefore, assuming a diameter of 1~arcsec for the emitting region (corresponding to the area of the 5 pixels in the black cross in Fig.~\ref{fig16}) used to generate the DEM, we estimate an emitting volume of 3.28$\times$10$^{22}$~cm$^{3}$. As a result, the electron density at the loop-top source in the TR plasma varies between 7.45$\times$10$^{9}$~cm$^{-3}$ and 1.10$\times$10$^{10}$~cm$^{-3}$ comparing between changes in the profiles at 16:09:44~UT and 16:10:59~UT, i.e. around the time of the chromospheric rain formation. The densities that we calculate are representative of the EUV component of the plasma and we fully expect the eventual chromospheric component to have a much higher density \citep[as demonstrated in][]{2015ApJ...806...81A}. Assuming a momentary pressure balance in the loop-top source, before the onset of the multi-thermal flow along the loop legs away the loop-top at frame 16:10:59~UT (see Fig.~\ref{fig10b}), we can estimate the density of the H$\alpha$ component with gas temperature of $\sim$22,000~K from Fig.~\ref{fig6}. Equating the ideal gas law, we estimate the density of the chromospheric component of coronal rain in this time range to be 9.21$\times$10$^{11}$~$\pm$1.76$\times$10$^{11}$~cm$^{-3}$. In the chromospheric component of the coronal rain, we detect a range of cross-sectional widths of rain clumps in the range of 100-200~km within H$\alpha$ images, across the EUV loop (refer to \citet{2014ApJ...797...36S} for more information) amounting to 8 parallel strands. Using the rain strand widths, as a lower limit to the volume of the cooling strands, together with the electron densities estimated here from chromospheric plasma, we calculate the mass loss rate from the post-flare arcade to be as much as 1.98$\times$10$^{12}$~$\pm$4.95$\times$10$^{11}$~g~s$^{-1}$.

\par As the plasma continues to cool below 1-2~MK in flare loops one must consider the shape of the optically thin loss function. When we consider again the radiative loss timescale using the \citet{1978ApJ...220..643R} radiative loss scaling law (for optically thin plasma) in the range of $10^{5.7}-10^{6.3}$~K (i.e. the lowest temperature range below which the plasma is assumed to be optically thick) we change the expression for the radiative cooling timescale to

\begin{equation}
\tau_{r}\,=\,3\,\times\,10^{21.94}\,\frac{k_{B}\,T}{n_{e}}.
\label{radiative_loss2}
\end{equation}

\par The near-simultaneous and parallel-forming rain clumps might be explained by a very short radiative loss timescale in individual strands, arising from modifications to the scaling law in the different temperature regimes, resulting in a more rapid temperature decline with relatively small change in density. For the derived TR plasma densities at the time of chromospheric rain formation and a plasma temperature of Log$_{10}$~T~=~5.9, which matches the expected temperature from the red curve of Fig.~\ref{fig18}, we calculate a new radiative cooling timescale of 2-3~min which continues to be greater than the expected / observed timescale. So far we have attempted to characterise the flare loop cooling processes for this event using simplified model assumptions which have largely proved very accurate, although, the model consistently predicts longer total loop cooling times than observed in this flare. 


\begin{figure}[!ht]
\centering
\includegraphics[clip=true,trim=0cm 2cm 0cm 3cm, scale=0.3, angle=0]{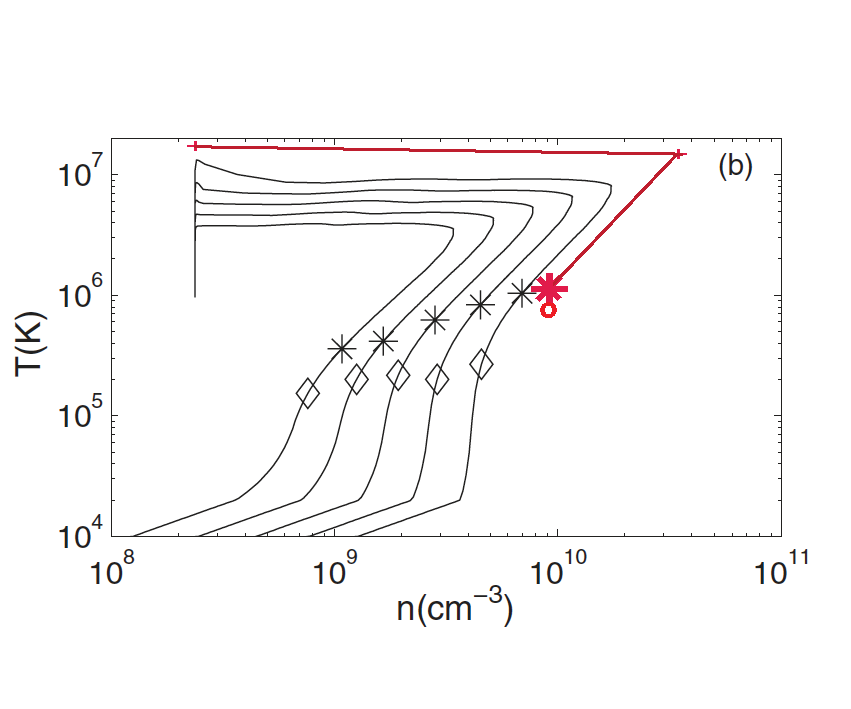}
\caption{This figure is extracted from \citet{2013ApJ...772...40C} (Fig.~1b) and the black curves describe the relation between the average temperature and density for 5 loops that are previously discussed as numerical simulation runs 6-10 in \citet{2010ApJ...717..163B}. Time increases when each curve is followed in a clockwise direction. The group of 5 loops have lengths (2$L$) in the range 67-72~Mm.  The black stars and diamonds show the start and end of the transition to catastrophic cooling according to the model. The red star corresponds to the temperature and density at the start of catastrophic cooling from the observations, to compare with the corresponding simulated black stars. The red crosses marks the temperature and density at the EM / density peak from the observations, to compare with the corresponding density peak of the simulated black curves. The red circle corresponds the the analytical solution for the critical temperature for the onset of catastrophic cooling ($T_{c}$) from Equation~\ref{catcool2}.}
\label{fig21}
\end{figure}


\par \citet{2012A&A...543A..90R} modelled radiative loss functions derived from the recent CHIANTI database \citep{2012ApJ...744...99L}. They found relatively 'faster' onset of catastrophic cooling in loops, thereby reducing the total loop cooling time, resulting in the plasma temperature dropping below 10$^{5}$~K in tens of seconds. This is thought to be attributed to a factor of 4 increase in the coronal radiative losses in the latest model when compared with \citet{1978ApJ...220..643R}. Furthermore, the flare loop temperature and density properties at the start of the radiative cooling phase described here are highly comparable with Case 1 from \citet{2012A&A...543A..90R}, as presented in \citet{2013ApJ...772...40C} Table~1, which suggests that the critical temperature for catastrophic cooling will be in the range of 10$^{6.0}$-10$^{6.2}$~K. As mentioned previously, we expect the start of catastrophic cooling occurs in the TR / low corona plasma with at least 10$^{5.9}$-10$^{6.2}$~K in temperature. Furthermore, we have shown that an earlier onset of catastrophic cooling is taking place and it appears within the range of 35-124~s after the plasma has cooled into the 1.5-7~MK coronal temperatures. 

\par Numerical models describing the cycle of cooling of flare loops can be represented using $n_{e}$--$T$ diagrams \citep{1992A&A...253..269J}. The density vs. temperature variations for a group of 5 loops was originally presented in \citet{2010ApJ...717..163B} ({\it Table 1 runs 6-10}) and extracted from \citet{2013ApJ...772...40C} ({\it Fig.~1b}), are now presented here in Fig.~\ref{fig21}. We present this figure in order to place our measurements in the context of flare models that accurately take into account the late process of catastrophic cooling and loop depletion. In Fig.~\ref{fig21}, the temperature and density ranges estimated from the observations, corresponding to the loop-top rain source at the onset of catastrophic cooling are marked with a red star. Likewise, the temperature and density ranges estimated for the start of the radiative cooling phase~{\bf IV} are marked with a red cross. The loop lengths for the simulation runs (black curves) increase from left to right but vary by as much as 6\% and are comparable with the loop length from our observations. Comparing the runs from right to left, each loop has a sequentially larger volumetric heating rate and so the radiative cooling starts with sequentially higher temperatures and densities. The volumetric heating rate almost doubles for each loop, increasing from 5.31$\times$10$^{-3}$~erg~cm$^{-2}$~s$^{-1}$ (left most curve) to 8.5$\times$10$^{-3}$~erg~cm$^{-2}$~s$^{-1}$ (right most curve).  It is clear to see that the observations presented (directly connected by the red line) may well represent the next 'class' of loop in this particular grouping, which would correspond to a loop which has been volumetrically heated (during phase~{\bf I}), by as much as 17.0$\times$10$^{-3}$~erg~cm$^{-2}$~s$^{-1}$, i.e. if the trend is indeed linear. 

\par We also find a strong agreement between the starting temperature of the simulated catastrophic cooling phase (marked by the black stars) and the observed starting temperature range of catastrophic cooling from the observations (red star). The red circle in Fig.~\ref{fig21} marks the analytical solution for the critical temperature defining the commencement of catastrophic cooling, which is again in very good agreement with the observations here, as it is with the numerical models outlined in detail in \citet{2013ApJ...772...40C}. What this means is that the onset of catastrophic cooling can indeed begin in the coronal plasma and very quickly accelerate through the transition region passbands to chromospheric temperatures without a significant change in density. Furthermore, the previous discrepancy between the analytical total loop cooling time and observed total loop cooling time, is now more closely matched and this discrepancy is also acknowledged in \citet{2013ApJ...772...40C}. Importantly, in the case of relatively short, hot flaring loops the models predicts that we can expect catastrophic cooling to commence at $\sim$1~MK, as our observations indicate. With a valid model comparison with our observations, how can the model inform the observations regarding the origin of catastrophic cooling leading to coronal rain at the loop-top?

\par It is considered that after the plasma is evaporated from the chromosphere, along each loop leg, the resulting compression near the loop top could generate slow-mode acoustic waves. In \citet{2013ApJ...772...40C} the definition for the critical temperature describing the onset of catastrophic cooling suggests an important role in the propagation of such sound waves in cooling loops and this process manifests itself in the ratio between the sound travel time and radiative cooling time. These sound waves may transport and redistribute a substantial amount of energy and dictate the relative importance of pure radiative cooling over enthalpy-based radiative cooling which balances the losses and sustains cooling. \citet{2013ApJ...772...40C} have shown that when the $T\,\sim\,n$ scaling in the loop cooling starts to break down then the temperature at which this happens defines is the onset of catastrophic cooling. A new scaling dependency appears as $T\,\sim\,n^{\delta}$, where $\delta$ is the typically 2 for short loops but can reduce to 1 for long loops and is determined by the relative importance of the coronal radiative losses to the enthalpy flux \citep{2010ApJ...717..163B}. Larger $\delta$ values indicates the dominance of radiation with small coronal mass loss, whereas smaller values indicate the dominance of enthalpy and a relatively large coronal mass loss \citep{2005A&A...437..311B}. \citet{2013ApJ...772...40C} state that the downflow required by enthalpy flux continually adjusts through sound waves that will sufficiently maintain the $T$\,$\sim$\,$n$ relationship provided the radiative cooling time in the corona ($\tau_{r}$) is greater than the sound travel time ($\tau_s$). Since sound wave travel time is determined by the loop length and sound speed, so loop length and loop temperature will also limit the expected onset of catastrophic cooling, aside from the volumetric heating rate of the flare. The sound travel time is therefore defined as

\begin{equation}
\tau_{s}\,=\frac{L}{C_{s}} 
\label{catcool}
\end{equation}

\par where the isothermal sound speed, $C_{s}$ = (2$k_{B}$T/m$_{p}$)$^{1/2}$ for an electron-proton plasma and m$_{p}$ is the proton mass. When $\tau_r~\le~\tau_s$, enthalpy-based radiative cooling stops and the loop cools predominantly by radiation, leading to catastrophic cooling. After equating these timescales and taking the conditions describing the start of the radiative cooling phase with subscript "0" (i.e. using $\tau_{r0}$ and $\tau_{s0}$ = 68~s), \citet{2013ApJ...772...40C} derived a general expression for the critical temperature ($T_{c}$) for the onset of catastrophic cooling as

\begin{equation}
T_{c} = T_{0}(\tau_{s0} / \tau_{r0})^{\frac{3}{2}-\delta-\frac{1}{\alpha}}.
\label{catcool2}
\end{equation}

\par From \citet{2010ApJ...717..163B} it is shown in Table~2 that for simulation runs 6-10 (corresponding to loops right to left in Fig.~\ref{fig21}) the value for $\delta$ generally decreases relatively linearly from 1.94 to 1.79. If the observations assigned to Fig.~\ref{fig21} indeed correspond to the next "class" of model parameters. Then given the parallels between the red line connecting the red symbols and the adjacent black curves, we could expect that the value for $\delta$ might equally scale up, so should be greater than 1.94 and very close to 2.0. Therefore, from the conditions at the loop-top source at the end of phase~{\bf IV}, with large $\delta$ will expect radiative cooling to dominant over enthalpy-based radiative cooling. After substituting into this expression our estimations for the temperature, radiative cooling timescale, sound travel timescale and with $\alpha$=-1/2 we determined the critical temperature of 0.78~MK (i.e 10$^{5.8}$~K). This is marked in Fig.~\ref{fig21} with the red circle and is in good agreement with the onset of steep gradients in the temperature profile described by the red curve in Fig.~\ref{fig18} at the transition to phase~{\bf V}, i.e 10$^{5.9}$-10$^{6.2}$~K. 

\par This analysis suggests that at the loop apex, where we do not expect strong flows (compared with loop legs) the radiation cooling should dominant and the model suggests it does. This leads to a run away cooling process for hot short loops (with a relatively short sound travel timescale), leading to catastrophic cooling at temperatures of around $\sim$1-1.5~MK in flares, which is in close agreement with these observations. If the sound waves cannot sustain the coronal radiative losses then catastrophic cooling is initiated and quickly the local temperature drops to chromospheric levels within 10s of seconds and we have rain flows in H$\alpha$. One interesting correlation is the similarity between $\tau_s$ which is 68~s and the periodicity in the H$\alpha$ rain flows of 55-70~s, along a given loop trajectory (see Fig.~\ref{fig7}). On this point, one hypothesis is that possibly the incident rain flows from the loop-top are triggering plasma compressions between the substrands, i.e. compressions within the coronal / TR medium of the loop arcade system, triggering new sound / acoustic wave propagation in their wake. At which point, the waves continue to replenish the energy balance requirements of the coronal and TR radiative losses, for another 68~s, until the condition for catastrophic cooling is met again, leading to further pressure balancing, accretion and loop drainage in new H$\alpha$ rain clump. That cyclic behaviour between the dominance of radiative vs. enthalpy-based radiative cooling might account for the sequential formation of rain flows in the loop system. Therefore, one might expect a linear relationship between sequential rain formation period and loop length at constant density. Furthermore, strong acoustic waves have been predicted for short heat pulses in flare loop models on the period of a few minutes \citep{2016ApJ...826L..20R}, which might account for the fluctuations in the GOES emission measure, as shown in Fig.~\ref{fig18}.
\newline

\subsection{Fine-scale structure in coronal loops}

\par The Cargill model accounts for this acceleration in the loop cooling time in one such loop strand (with 0D assumptions). However, we observe multiple, parallel strands of rain flows within the loop system. If successive chromospheric rain strands can form in parallel, each independently cooling catastrophically, then this model will be well suited to studying rain formation in post-flare arcades, collectively. This scenario may be plausible given that we can already identify multiple, independent coronal rain formation sites at the loop-top from the H$\alpha$ temperature maps (see Fig.~\ref{fig6}). Each of these may correspond to a unique strand within the loop system independently cooling and yielding rain flows.  \citet{2012ApJ...746...18R} showed that if a bundle of sub-loops (strands in this case) is assumed to be heated at slightly different times then they could obtain a closer match between the observed and modelled light curves for plasma below 2~MK in flare arcades. A future statistical study of the individual rain strands formed throughout this loop-system and their associated DEM properties, may yield further insight into the possibility of a collective convolution of coolings and the net effect on the estimated timescales for cooling may adjust slightly. Indeed, we find some evidence that chromospheric rain clumps in the loop arcade can form at slightly different times and they can even appear periodically (see Fig.~\ref{fig7}), indicating some underlying physics connecting the formation of successive, parallel clumps of rain. The multi-stranded nature of coronal rain may not be entirely thermally isolated and there could be energy transfer between adjacent strands as a result of sympathetic cooling which has been demonstrated numerically in the case of quiescent coronal rain formation, leading to parallel rain strand formation \citep{2013ApJ...771L..29F}. The transport of thermal energy via the MHD thermal mode (also known as the entropy mode in the absence of thermal conduction) was first predicted by \citet{1965ApJ...142..531F} between parallel strands cannot be ruled out in having a role in accelerating the cooling process. The generation of this thermal wave is guaranteed by the small but non-zero perpendicular thermal conduction in the corona. The spatial distribution of such mode, across a set of magnetic field lines, results in a set of dense rain clump formations with multiple smaller clumps located beside each other and of similar widths. This process will require further investigation with more sophisticated, multi-stranded models of cooling loops to determine the impact of the thermal mode with respect to the formation of adjacent coronal rain clumps \citep{2015ApJ...806...81A}. 

\par In general, it is most likely that the structure of the post-flare arcades is composed of a bunch or bunches of loop strands, rather than a single strand and the approximations here are based on single strand models. \citet{2010A&A...513L...2K}~\&~\citet{2010ApJ...717..250K} use RHESSI observations of sources of X-ray emission during flaring to show that a chromospheric density model involving multiple density threads can explain both the position of the maximum and the vertical size of the sources. Likewise, \citet{2012ApJ...746...18R} found a correlation with SoHO / SUMER \citep{1997SoPh..170...75W} spectral lines under flaring conditions when they assumed a model of flare heating within bundle of sub-loops of equal length. A multi-stranded model will be important to consider if we only focused on the GOES X-ray data and / or our analysis involved studying the loops in their entirety in the EUV and / or visible / near-IR wavelengths. If we wanted to draw conclusions about the nature of the loop in its entirety then we should consider the convolution of cooling of the individual strands in the bunch and the filling factor associated with flaring plasma in the arcades. On that point, we should not expect that all sub-strands of the post-flare loop system are 100\% filled with heated plasma, all at the same time, during the flare impulsive phase. Instead, here we focus on deducing plasma properties of the loop-top coronal rain formation region which composes part of a single chromospheric coronal rain strand within a small region of the entire multi-thermal loop-system (see Figs.~\ref{fig4},~\ref{fig10a}~\&~\ref{fig11}). In summary, the application of single strand model assumptions in our timescale analysis of the formation of sub-million Kelvin rain is valid.  

\par It is important to note that the cooling timescales calculated are only applicable to the linear regime and catastrophic cooling is a non-linear effect. We emphasise that the plasma properties determined here correspond to hotter components of the multi-thermal rain sources and not the final product i.e. not the chromospheric component of the rain which is optically thick. Indeed, continually increasing densities in the cooling plasma will also shorten the ionisation and recombination timescales and allow the plasma to efficiently adjust its ionisation status \citep{1989SoPh..122..245G}, i.e. the plasma should respond quickly to the rapidly changing conditions. Furthermore, it is important to state that we still need to confirm whether the observed radiative cooling rates can continue down the low (sub-million Kelvin) temperature bins, given the rates of condensation leading to optically thick chromospheric plasma, since the reduced efficiency of radiative losses within optically thick plasma in a coronal loop-top environment may not account for such a rapid cooling timescale alone. Hence, based on this study we do not know whether or not we still need extra mechanisms, such as the MHD thermal mode, in order to explain catastrophic cooling into the chromospheric component of the rain in flaring conditions or in the case of quiescent coronal rain. In considering these effects on the measurements deduced in this study, together with the inherent uncertainties in the measurement of densities from observation of the DEMs, leads us to reiterate that the reported timescales and the expected density enhancement in the chromospheric component of the flare-driven coronal rain will vary by a few factors.

\section{CONCLUSIONS}

\par We investigate in detail and at the highest resolution a "textbook" example of the flare loop cooling process from the start of the decay at a temperature peak of 15.4~MK in the X-ray channels with GOES, through the EUV channels of the EUV corona and TR with AIA, finishing as mass condensations at chromospheric temperatures in the visible and near-IR with CRISP.  In doing so, we reveal the chromosphere-corona mass cycle from source, as bright H$\alpha$ ribbons to sink, as coronal rain. We detect 5 phases that characterise the post-flare loop dynamics: heating, evaporation, conductive cooling dominance for 122~s, radiative / enthalpy-based cooling dominance ($<$4700~s) and finally catastrophic cooling at a critical plasma temperature close to 10$^{5.8}$~K. In summary, we find an excellent agreement between the observations and analytical model predictions in all phases, primarily derived in \citet{1995ApJ...439.1034C} and \citet{2013ApJ...772...40C}, which very much agree with respect to the timescales associated with the dominant cooling processes, as well as, the critical temperature for the onset of catastrophic cooling. 

\par We discover that during catastrophic cooling of phase~{\bf V} the plasma cools at a max. rate of 22700~K~s$^{-1}$ in multiple loop-top sources and this presents itself as a catastrophic loop depletion in up to 8 parallel strands. If successive chromospheric rain strands can form in parallel, each independently cooling catastrophically, then the Cargill model will be well suited to studying rain formation in multi-stranded post-flare arcades. The acceleration in the rate of cooling to chromospheric temperatures is evidence for a catastrophic cooling process. The plasma can undergo catastrophic cooling from $\sim$1~MK to $\sim$22,000~K in tens of seconds (specifically 35-124~s here) rather than in many minutes as with quiescent coronal rain which is remarkably fast. This could be explained by the fact that the flaring process leads to substantially larger mass loading of the coronal loop system leading to shorter radiative cooling timescales.

\par We study the initial rain formation region at the loop apex in detail and we find strong evidence to suggest the presence of a multi-thermal flow within flare-driven coronal rain strands. From spatial correlation and tracking of the flows from loop-top along the legs, between the emitting regions in the EUV and chromospheric channels, we could identify the multi-thermal structure of the rain flows. Then with a DEM analysis, employing a novel regularised inversion code, we calculated the density of the EUV plasma from the emitting volume (estimated from the FOV). Assuming a pressure balance across the multi-thermal structure of the rain clump at the formation of the rain source at the loop-top, we estimated the density of the chromospheric component of the rain to be 9.21$\times$10$^{11}$~$\pm$1.76$\times$10$^{11}$~cm$^{-3}$ which is characteristic of quiescent coronal rain densities, as reported by \citet{2015ApJ...806...81A}. Using the rain strand widths, in the range of 100-200~km in cross-section, together with the electron densities estimated here from chromospheric plasma, we calculate the mass loss rate from the post-flare arcade to be as much as 1.98$\times$10$^{12}$~$\pm$4.95$\times$10$^{11}$~g~s$^{-1}$.

\par We detect catastrophic cooling at the loop-top source initially, followed by multi-thermal rain flows along the loop-leg leading to a notable increase in the TR emission measure at 10$^{5.8}$~K from the DEM maps for the loop-leg. Model predictions for the loop-top temperature and density properties ascribe to a down-flow velocity in the loop-leg with the same magnitude that is observed (i.e. 54.5~$\pm$2~\kms), suggesting a redistribution of the energy balance in the loop during the flow and a dominance of the enthalpy-flux in powering the TR emission. We also detect a deceleration in the rain flow closer to the loop footpoint from 76-60~m~s$^{-2}$ which is consistent with numerical simulation of rain flows into a more dense medium of the TR and chromosphere.The role of enthalpy flux in sustaining the TR losses may play a similar role in sustaining the corona losses and the balance between enthalpy-based radiative cooling and unsustained radiative cooling dictates the onset of catastrophic cooling. What drives the transition to catastrophic cooling is the sound wave travel time according to \citet{1995ApJ...439.1034C}. We reveal a close proximity between the model predictions (of 10$^{5.8}$~K) and the observed properties (between 10$^{5.9}$~K and 10$^{6.2}$~K) defining the temperature onset of catastrophic cooling. This suggests that the role of sound waves in loops, through sustaining the enthalpy flux, may be important when determining the eventual onset of H$\alpha$ rain formation. 

\par Finally, we postulate that the ratio between radiative cooling timescales and sound wave travel timescale may not only be important in calculating the critical temperature for the onset of catastrophic cooling and coronal rain formation in flares but in explaining the origin of the periodicity between rain clumps which is 68~s and is on the order of the sound wave travel time calculated here. Future work will focus on understanding the energy balance between interacting loop strands collectively and the physics associated with the formation of parallel adjacent rain flows in post-flare arcades.

\acknowledgements

The authors are most grateful to the staff of the SST for their invaluable support with the observations. The Swedish 1-m Solar Telescope is operated on the island of La Palma by the Institute for Solar Physics at Stockholm University in the Spanish Observatorio del Roque de los Muchachos of the Instituto de Astrof{\'\i}sica de Canarias.  ES is a Government of Ireland Post-doctoral Research Fellow supported by the Irish Research Council. ES would like to acknowledge the DJEI/DES/SFI/HEA Irish Centre for High-End Computing (ICHEC) for the provision of computational facilities (in particular FIONN cluster) and support. EPK was supported by STFC. ES would like to acknowledgement the support from the International Space Science Institute, Bern, Switzerland and to the International Team involved in the project, "Implications for coronal heating and magnetic fields from coronal rain observations and modeling", lead by PA. GV is funded by the European Research Council under the European Unions Seventh Framework Programme (FP7/2007-2013)\,/\,ERC grant agreement nr.~291058. SW acknowledges support by the Research Council of Norway (grant 221767/F20).

\bibliography{bib}

\bibliographystyle{aa}

\end{document}